%% file: aanda.tex
\documentclass{aa}

\usepackage{marginnote}
\usepackage[dvipsnames]{xcolor}
\usepackage{graphicx}
\usepackage{float}
\usepackage{txfonts}
\usepackage{soul}
\usepackage{ulem}
\usepackage{booktabs}
\usepackage{subcaption}
\usepackage{amsfonts}
\usepackage{dcolumn}
\usepackage{tablefootnote}
\usepackage{threeparttable}
\usepackage{lsstdescmacros}

\newcolumntype{d}[1]{D..{#1}}

\newcommand{\add}[1]{\textcolor{black}{#1}}
\newcommand{\addtwo}[1]{\textcolor{black}{#1}}

\newcommand{\sqdeg}[1]{{\rm deg}$^2$}
\newcommand{\area}[1]{15~deg$^2$}
\newcommand{\areatot}[1]{300~deg$^2$}
\newcommand{\diasrc}[1]{\texttt{diaSource}}
\newcommand{\diasrcs}[1]{\texttt{diaSources}}
\newcommand{\diaobj}[1]{\texttt{diaObject}}
\newcommand{\diaobjs}[1]{\texttt{diaObjects}}
\newcommand{\snana}{{\fontfamily{qcr}\selectfont{SNANA}}}
\newcommand{\diapipe}{\texttt{dia\_pipe}}
\newcommand{\cosmoDC}{\texttt{cosmoDC2}}

\newcommand{\OM}{\Omega_M}

\newcommand{\sigint}{\sigma_{\rm int}}

\newcommand{\NzBBC}{9}  
\newcommand{\NimageTemplate}{40}

\newcommand{\PSF}{PSF}
\newcommand{\maglim}{m_{5\sigma}}
\newcommand{\matchtol}{0\arcsec.5}

\newcommand{\NmatchSRC}{17,719}
\newcommand{\NmatchSN}{2186}
\newcommand{\mhalf}{m_{1/2}}
\newcommand{\SNRhalf}{\rm{SNR}_{1/2}}
\newcommand{\eff}{efficiency}
\newcommand{\effsym}{\epsilon}
\newcommand{\effSN}{\epsilon_{\rm SN}}
\newcommand{\Ftrue}{F_{\rm true}}
\newcommand{\sigF}{\sigma_F}
\newcommand{\SKYNAME}{DC2-SNIa}

\newcommand{\Bd}{D_{\rm art}}
\newcommand{\Ndiasrc}{N_{\texttt{diaSrc}}}
\newcommand{\NsrcSN}{\Ndiasrc{}^{\rm{SN}}}
\newcommand{\Nsrcart}{\Ndiasrc{}^{\rm{art}}}
\newcommand{\Nsrcvar}{\Ndiasrc{}^{\rm{var}}}

\newcommand{\mSB}{m_{\rm SB}}
\newcommand{\mSBcompare}{\mSB(\RMSpull{=}2)}

\newcommand{\RMSpull}{{\rm RMS}_{\rm pull}}

\newcommand{\nameSimData}{DATA-like}
\newcommand{\LowZName}{SimLow-z}
\newcommand{\NLSSTCUTS}{504}
\newcommand{\NLOWZCUTS}{151}
\newcommand{\NDATACUTS}{655} 
\newcommand{\NSIMLOWZCUTS}{501}
\newcommand{\NSIMDATACUTS}{1560} 
\newcommand{\NSIMCUTS}{2061} 
\newcommand{\NBIASCOR}{6.8\times10^5}

\newcommand{\zhalfsym}{z_{1/2}}
\newcommand{\zhalfDetect}{0.730}
\newcommand{\zhalfCuts}{0.58}
\newcommand{\zhalfDetectSim}{0.730}
\newcommand{\zhalfCutsSim}{0.61}
\newcommand{\Rcmb}{R(z_{*})}
\newcommand{\sigR}{\sigma_{R}}
\newcommand{\sigRvalue}{0.007}
\newcommand{\mtrue}{m_{\rm true}}
\begin{document}


   \title{SNIa-Cosmology Analysis Results from Simulated LSST Images: from Difference Imaging to Constraints on Dark Energy}

    \author{
B.~S\'anchez \inst{1},
R.~Kessler \inst{2,3},
D.~Scolnic \inst{1},
B.~Armstrong \inst{4},
R.~Biswas \inst{5},
J.~Bogart \inst{6},
J.~Chiang \inst{6, 7},
J.~Cohen-Tanugi \inst{8, 10},
D.~Fouchez \inst{9},
Ph.~Gris \inst{10},
K.~Heitmann \inst{11},
R.~Hlo\v{z}ek \inst{12},
S.~Jha \inst{13},
H.~Kelly \inst{6, 7},
S.~Liu \inst{14, 15},
G.~Narayan \inst{16, 17},
B.~Racine \inst{9},
E.~Rykoff \inst{6}
M.~Sullivan \inst{18},
C.~Walter \inst{1},
M.~Wood-Vasey \inst{14, 15}
\begin{center} The LSST Dark Energy Science Collaboration (DESC) \inst{ }\end{center}
}

\scriptsize
\institute{
Department of Physics, Duke University, Durham, NC 27708, USA
\and Department of Astronomy and Astrophysics, University of Chicago, Chicago, IL 60637, USA
\and Kavli Institute for Cosmological Physics, University of Chicago, Chicago, IL 60637, USA
\and Lawrence Livermore National Laboratory, 7000 East Ave., Livermore, CA 94550-9234, USA
\and Stockholm University, Universitetsv\"agen 10, 114 18 Stockholm, Sweden
\and SLAC National Accelerator Laboratory, 2575 Sand Hill Road, Menlo Park, CA, 94025, USA
\and Kavli Institute for Particle Astrophysics and Cosmology, Stanford University, Stanford CA 94305, USA
\and Laboratoire Univers et Particules de Montpellier, Place Eug\`ene Bataillon - CC 72, F-34095 Montpellier Cedex 05, France
\and Centre de Physique des Particules de Marseille, 163, avenue de Luminy Case 902 13288 Marseille cedex 09, France
\and LPC, IN2P3/CNRS, Université Clermont Auvergne, F-63000 Clermont-Ferrand, France
\and HEP Division, Argonne National Laboratory, 9700 S Cass Ave, Lemont, IL 60439, USA
\and University of Toronto, 27 King's College Cir, Toronto, ON M5S, Canada
\and Department of Physics and Astronomy, Rutgers University, Piscataway, NJ 08854, USA
\and Department of Physics and Astronomy, University of Pittsburgh, 4200 Fifth Ave, Pittsburgh, PA 15260, USA
\and Pittsburgh Particle Physics, Astrophysics, and Cosmology Center (PITT PACC).
\and Department of Astronomy, University of Illinois at Urbana-Champaign, Urbana, IL 61801, USA
\and Center for AstroPhysical Surveys (CAPS), National Center for Supercomputing Applications (NCSA), University of Illinois at Urbana-Champaign, Urbana IL 61801, USA
\and School of Physics and Astronomy, University of Southampton, University Rd, Southampton SO17 1BJ, United Kingdom.
}

    \normalsize
    \authorrunning{S\'anchez, Kessler, Scolnic \& the LSST DESC}
    \titlerunning{SNIa-Cosmology from Simulated LSST Images}


   \date{Received - ; accepted - }


 \abstract
   {
   The Vera Rubin Observatory Legacy Survey of Space and Time (LSST) is expected to process ${\sim}10^6$ transient detections per night.
   To use these transients for precision measurements of cosmological parameters and rates studies, it is critical to understand the detection efficiency, magnitude limits, artifact contamination levels, and biases in the selection and photometry.
   Here we rigorously test the LSST Difference Image Analysis (DIA) pipeline using simulated images from the Rubin Observatory LSST Dark Energy Science Collaboration (DESC) Data Challenge (DC2) simulation for the Wide-Fast-Deep (WFD) survey area.
   DC2 is the first large-scale (\areatot{}) image simulation of a transient survey that includes realistic cadence, variable observing conditions, and CCD image artifacts.
   We analyze ${\sim}$\area{} of DC2 over a 5-year time-span in which artificial point-sources from Type Ia Supernovae (SNIa) light curves have been overlaid onto the images.
   We measure the detection efficiency as a function of Signal-to-Noise Ratio (SNR) and find a $50\%$ efficiency at $\rm{SNR}=5.8$ averaged over all bands.
   The corresponding magnitude limits for each filter are: $u=23.66$, $g=24.69$, $r=24.06$, $i=23.45$, $z=22.54$, $y=21.62$ \magn.
   The artifact contamination levels is $\sim90\%$ of all detections, corresponding to $\sim1000$ artifacts/\sqdeg in $g$ band, and falling to 300/\sqdeg in $y$ band.
   The recovered photometry has biases $<1\%$ for magnitudes $19.5 < m <23$.
   We show that our DIA performance on simulated images is similar to that of the Dark Energy Survey difference-imaging pipeline applied to real images.
   We also characterize DC2 image properties to produce catalog-level simulations needed for distance bias corrections.
   We find good agreement between DC2 data and simulations for distributions of SNR, redshift, and fitted light-curve properties.
   Applying a realistic SNIa-cosmology analysis for redshifts $z<1$, we recover the input cosmology parameters to within statistical uncertainties.
   Finally, we discuss further applications of this dataset and analysis, and we suggest pipeline improvements before LSST operations begins.
   }

   \maketitle
%

\section{Introduction}
\label{sec:intro}
The Vera C. Rubin Observatory Legacy Survey of Space and Time (LSST\footnote{\url{http://www.lsst.org}}, \citealp{Ivezic2019}) is expected to soon begin operations and acquire images over 10 years.
This survey will use the \add{Simonyi Survey Telescope at }Rubin Observatory, which is a 8.4m class\footnote{6.7m of effective collecting area} telescope with a 3.2 Gigapixel camera, yielding a 9.6 deg${}^2$ field of view.
The \add{Rubin Observatory LSST Camera} design includes \textit{ugrizy} filters, and the expected $5\sigma$ $r$~band depth is \add{$>24$} (AB system) in a single 30 second visit, where each visit is comprised of two 15 second exposures.
The instrument and the survey strategy have been optimized towards obtaining repeated observation of ${\sim} 20,000$~\sqdeg{} of the sky over 10 years.

LSST will explore a broad range of research fields in astrophysics \citep{LSSTScienceCollaboration2009}; the main science objectives are the study of solar system dynamics, mapping the Milky Way structure, and probing dark matter and dark energy.
\add{Many of} these science goals rely on the discovery of transient sources, and the expected number of transient detections from all astrophysical variability sources is ${\sim}10^6$ per night\footnote{\url{https://www.lsst.org/scientists/keynumbers}} \citep{Ivezic2019, Ridgway2014, Graham2020}, an unprecedented rate when comparing to precursor surveys.
Past transient surveys have either focused on low redshift \add{($z<0.1$)} using shallow/wide area strategies, or higher redshift using deep/limited area strategies (see \citealp{Scolnic18} for a review).
The unique capabilities of LSST enable survey strategies using wide areas with deep images.
To discover transients \add{and measure their light curves, the LSST Project has developed Difference Image Analysis (DIA) software components}.
\add{The Rubin Observatory} LSST Dark Energy Science Collaboration (DESC\footnote{\url{http://lsstdesc.org}}) used these components to develop an orchestration software layer called  \diapipe{}\footnote{\url{https://github.com/LSSTDESC/dia_pipe}}.
In this paper, we make the first evaluation of this pipeline by analyzing simulated images.

Type Ia supernovae (SNIa) are transient events that are used as cosmological probes to measure the expansion history of the universe and in particular the dark energy equation of state $w$ (and its cosmic evolution parameterized by $w_a$; \citealp[SRM]{DESCCollab2019_SRM}).
LSST is expected to increase the SNIa sample size by up to a factor of 100 compared to previous samples \citep{betoule_jla_2014, sako_sdssSn_2018, Scolnic18, jones_foundation_2019}.
Furthermore, the survey will yield the discovery of SNIa using a single instrument with redshifts up to $z \sim 1.2$.
The requirements on systematic uncertainties from the SNIa-cosmology analysis are detailed in \cite{mandelbaum_srd_2018};
\add{these requirements include}
photometric precision at the few mmag-level and accurately determined selection biases.

To rigorously test analysis pipelines before the pre-survey commissioning period, DESC has generated a large and comprehensive set of image simulations known as Data Challenge 2 (DC2) \citep{sanchez_DC1_2020, desc_dc2_2020}.
Each DC2 image is based on models of the LSST instrument \add{and expected observing conditions at the summit}
(sky noise and Point Spread Function - PSF),
\add{along with} realistic catalogs of galaxies and supernovae light-curves.
The full DC2 area covers \areatot{} of the Wide-Fast-Deep (WFD) survey and includes injected point sources of Type Ia supernovae with an average cadence of $3$ days ($15$ days in each filter).
Using a DC2 subset of \area{}, we have processed the raw CCD pixels with DIA to characterize SNIa transient finding, photometric precision, and selection effects.
In addition, we treat DC2 like real data and perform a cosmology analysis that includes light-curve fitting, bias-correcting distances, and fitting for cosmological parameters $w$ and $\OM$. This pixel-to-cosmology test is a critical part of evaluating \diapipe{} readiness for survey operations.

The layout of this work is the following. In Section \ref{section:DC2} we explain the DC2 dataset used and in Section \ref{section:analysis} we give details on the analysis and the techniques implemented.
Section \ref{section:dataresults} shows our results on DC2 data processing, and Section \ref{section:metrics} lists performance metric scores obtained for transient detection as well as cosmology fitting.
In Section \ref{section:discussion} we discuss our results and compare them to previously reported analysis.
The final discussion and conclusions are presented in Section \ref{section:conclusions}.

%
\section{The DC2 Dataset}
\label{section:DC2}
\add{The Data Challenge 2 (DC2)} is a \add{broad} DESC effort to create \add{and process} simulated LSST images based on modelling galaxies and transients in the universe \citep{desc_dc2_2020}. The simulation is composed of observations spanning a sky area of 300~\sqdeg{} during 5 years of survey operations. The simulated data includes the expected instrumental signatures from the LSST Camera as well as the atmospheric effects in all six optical bands \textit{ugrizy}. DC2 contains stars, galaxies, and astrophysical effects such as clustering, \add{cosmic web/}structure formation, and gravitational lensing effects such as cosmic shear.
DC2 also includes \add{variable stars, transient variability from} SNe~Ia\footnote{No other SN types are included}, \add{Active Galactic Nuclei (AGN) galaxies}, and strong lensed SNe~Ia.

DC2 used the state-of-the-art N-body simulation \textit{Outer Rim} (\citealp{Heitmann2019}). The cosmological parameters used to create \textit{Outer Rim} are consistent with \textit{WMAP-7} \citep{komatsu_7yrWMAP_cosmo}.
From this gravity-only simulation, the \cosmoDC mock catalog (\citealp{korytov_cosmoDC2_2019}) \add{is} created;
it covers 440~\sqdeg{} of sky area up to a redshift of $z=3$.
CosmoDC2 contains more than 500 properties for each galaxy, including stellar and halo mass, shape,
\add{Spectral Energy Density (}SED), central black hole parameters, AGN activity, as well as environment related quantities such as the full gravitational shear and convergence maps of the sky,
which gives the observed shape of each galaxy.

In this work we use Run~2.2i Wide Fast Deep (WFD) images with an average transient cadence of 3 days between observations. DC2 images were produced with the image simulation software \texttt{imSim}\footnote{\url{https://github.com/LSSTDESC/imSim}} that imprints observing conditions and instrumental signatures using a model of the LSSTCam. The observing conditions, which include sky noise, Point Spread Function (PSF), zeropoint, and dithering are based on the \texttt{minion\_1016}\footnote{\url{http://ls.st/Collection-4604}} observing strategy produced with the operations simulator software \texttt{OpSim}\footnote{\url{https://github.com/lsst/sims_operations}}. Each object SED is attenuated from a calculation of Galactic dust extinction and atmospheric effects that includes differential chromatic refraction. Effects from the CCD readout electronics are also simulated, including charge repulsion effects and saturation. Each simulated visit illuminates 189 $4k\times4k$ CCD detectors (3 billion pixels) covering almost 10 \sqdeg{}, with a plate scale of 0.2 \asec/px.

A summary of the DC2 SNIa properties is shown in Table~\ref{tab:salt2dc2pars}. For SNIa, the rest frame SED is computed with the SALT2 model \citep{guy2010supernova, betoule_jla_2014}. Since the original SED model covers only the $g$ and $r$ bands in the rest-frame, we use a wavelength-extended model \citep{pierel_extending_2018}  that covers all of the LSST bands. For \texttt{imSim} to run properly, we include an additional modification that prevents negative UV spectral fluxes. The properties of each SNIa in DC2 are determined by the following SALT2 parameters: redshift ($z$), time at peak brightness ($t_0$), stretch ($x_1$), color ($c$), and amplitude ($x_0$). Each redshift is randomly selected from a volumetric rate, $r_v(z) = 2.5 \times 10^{-5}(1+z)^{1.5}\rm{Mpc}^3\rm{yr}^{-1}$ \citep{Dilday_2008}. The $t_0$ value is randomly selected within the 5 year DC2 time span.
Each SNIa includes an intrinsic scatter drawn from a Gaussian distribution with $\sigma=0.15$~\magn{}; a coherent~\magn{} fluctuation is applied at all SNIa phases and wavelengths.
The SALT2 parameters $x_1$ and $c$ were each drawn from an asymmetric Gaussian distribution with parameters shown in Table~\ref{tab:salt2dc2pars}. The amplitude $x_0$ is computed from the SALT2 parameters and the luminosity distance. The luminosity parameters $\alpha$ and $\beta$ were set to \add{$0.137$ and $3.21$} respectively.

\begin{table}[]
    \centering
    \caption{SNIa properties used in DC2 simulations.}
    \begin{threeparttable}
    \begin{tabular}{l|l}
       \toprule
        \multicolumn{2}{c}{DC2 SNIa property}  \\
        \midrule
        Light curve model            & SALT2 Extended + \add{0.15~mag offset\tnote{*}}\\
                                     & \citep{pierel_extending_2018} \\ \midrule
        Rate model                   & $r_v(z) = 2.5 \times 10^{-5}(1+z)^{1.5}\rm{Mpc}^3\rm{yr}^{-1}$ \\
                                     & \citep{Dilday_2008} \\ \midrule
        Intrinsic scatter            & \add{$\sigint = 0.15$\magn{}}  \\ \midrule
        Host correlation             & None \\ \midrule
        Stretch population           & $\overline{x}_1 = 0.873$,  \\
                                     & $\sigma_{+}=1.43$, $\sigma_{-}=0.359$,   \\
                                     & Range$=[-3.0, 2.0]$  \\ \midrule
        Color population             & $\overline{c} = -0.048$,   \\
                                     &  $\sigma_{+}=0.043$, $\sigma_{-}=0.097$,   \\
                                     &  Range$=[\pm0.3]$\\ \midrule
        Luminosity parameters        & \add{$\alpha=0.137$, $\beta=3.21$} \\
        \bottomrule
    \end{tabular}
\begin{tablenotes}\footnotesize
    \item[*]{\add{This offset was a mistake in the DC2 generation,
        and is included in the simulations for bias corrections.}}
\end{tablenotes}
\end{threeparttable}
    \label{tab:salt2dc2pars}
    \end{table}

Correlations between the SN~Ia and host-galaxy properties were not included in the simulations.
Supernovae were assigned to a galaxy using an occupation probability proxy of stellar mass.
For host galaxies of SNe~Ia, the \cosmoDC{} stellar mass distribution peaks at $\sim10^{11}\rm{M}_{\odot}$,
which is $\sim5$ dex higher than the typical mass of \add{field} galaxies
\add{(See Fig.~5 in \citep{desc_dc2_2020})}.
$10\%$ of SNe~Ia are assigned to be `hostless' in order to provide a control sample of isolated transients.

DC2 images are processed by the LSST pipelines, which organize image data in sets of \textit{tracts} and \textit{patches} for operational purposes.
Each 2.56~\sqdeg{} tract is a square containing $7\times7$ patches that share a common World Coordinate System (WCS) projection. This sky map organization is used for image-coaddition grid resampling, database access and image data manipulation, DIA processing, and template creation.
Each DC2 patch contains $4100\times4100$ pixels (roughly the size of a CCD) with a scale of $0.{\arcsec}2$. To avoid missing area due to edge effects, patches overlap with their
neighbors by 100 pixels and tracts overlap by 1~\amin.

\begin{figure}
    \includegraphics[width=\linewidth]{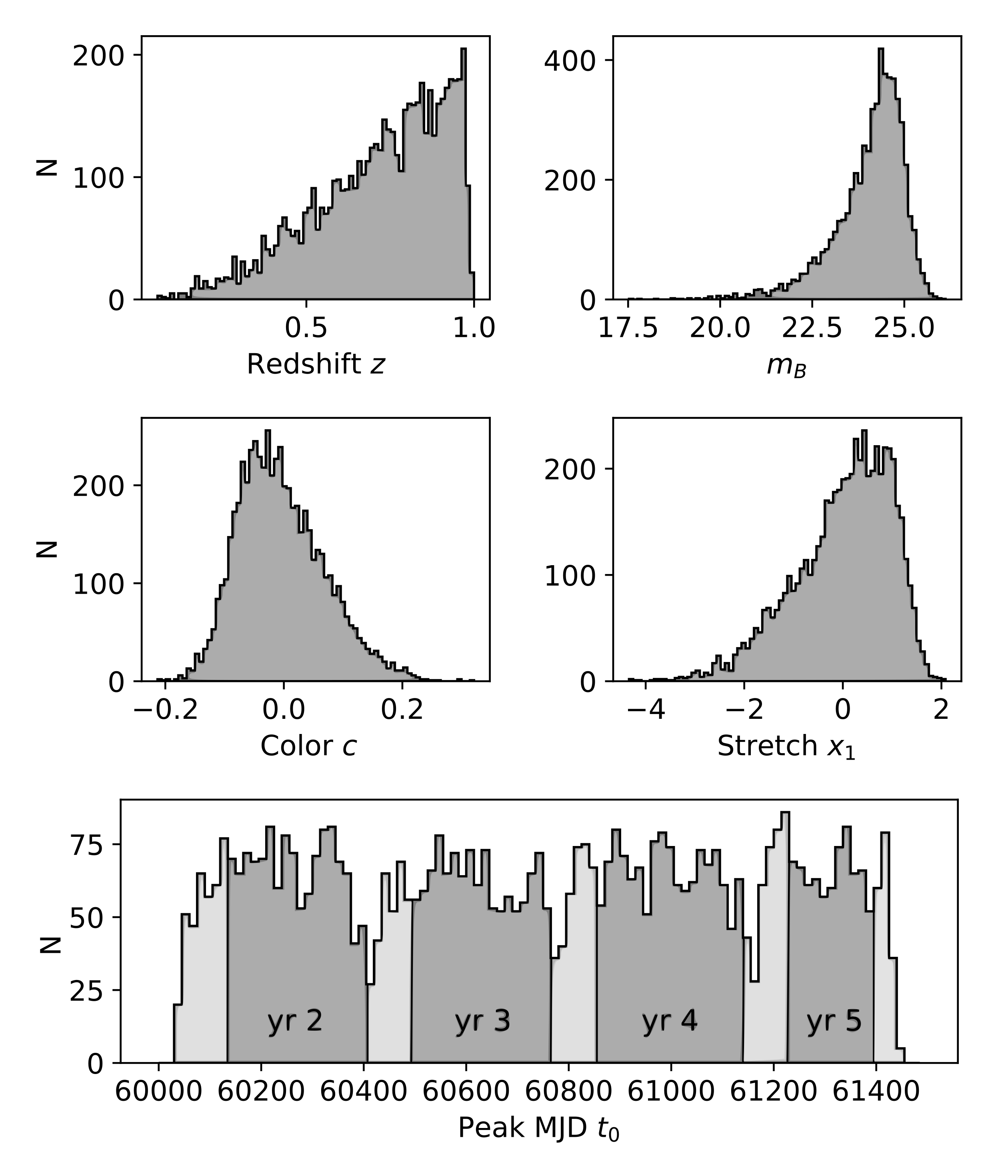}
    \caption{Distribution of redshift and SALT2 model parameters for
    DC2 Type Ia Supernovae. The light grey regions (bottom panel) are for events whose peak brightness
    occurs outside a season of observations.
    }
    \label{fig:dc2_salt2pars}
\end{figure}

\begin{figure*}
    \centering
    \includegraphics[width=\hsize]{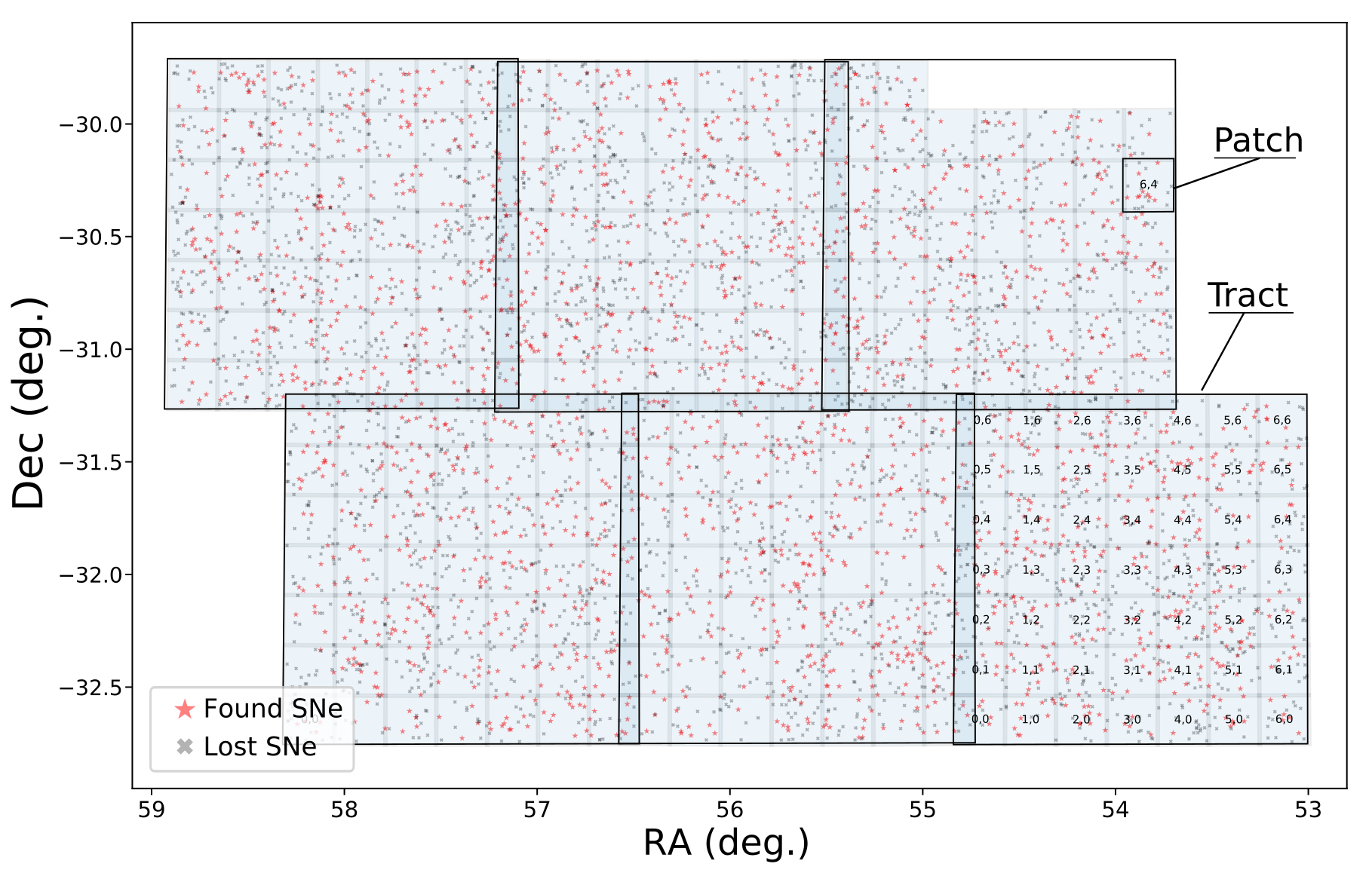}
       \caption{
           \SKYNAME{} sky area. Small squares indicate patches, and large squares are tracts. Detected and undetected SNe are shown as red stars and grey crosses, respectively. \add{In light-blue we show used patches.}
       }
    \label{fig:sky_area}
\end{figure*}

In this work, we select a \area{} area from the DC2 WFD (hereafter called ``\SKYNAME{}'' area) which includes 1967 LSSTcam visit observations and \(105,942\) CCD images.
We select SNe~Ia with redshifts $z\leq 1.0$, which includes the full range of cosmologically useful SNe Ia that will be discovered in the WFD \citep{mandelbaum_srd_2018}.
\add{W}e discard objects $<65\arcsec$ from the edges of the \SKYNAME{} area
\add{in order to avoid subtraction artifacts from template overlapping issues}.
\SKYNAME{} contains 5884 Type Ia SNe, and Fig.~\ref{fig:dc2_salt2pars} shows distributions of redshift and SALT2 parameters. Fig.~\ref{fig:sky_area} shows the \SKYNAME{} sky area used in this work, \add{illustrating } the tracts and patches \add{used}, and the locations of analyzed SNe~Ia.

%

\section{Analysis}
\label{section:analysis}
Here we describe the \textit{DIA LSST pipeline framework}, and SNIa-cosmological analysis.

\subsection{DIA pipeline framework}
\label{section:analysis-pipeframework}
The Rubin Observatory Data-Management team has developed a state-of-the-art set of
\add{software} tools for CCD data reduction that contains several routines for image processing, such as image coaddition, flux measurements, etc. This image processing framework, named ``LSST Science Pipelines''\footnote{\url{https://pipelines.lsst.io}}, is open source and can be used on any optical and infra-red survey data set.
DESC uses the LSST pipelines system to remove instrumental signatures from DC2 images,
(\add{e.g.,} electronic readout bias, dark current, illumination gradients), to calibrate images, and \add{to} obtain a World Coordinate System (WCS) solution.

For transient detection, DESC has developed a specific pipeline package \diapipe{} that uses LSST Science Pipelines' image processing tools, including Difference Image Analysis (DIA) routines. The central concept of DIA is to compare two images of the same sky area taken at different times, and detect sources that change in brightness. Each image has different properties (such as PSF, sky noise and zeropoint, etc.) and the subtraction accounts for these effects. DIA uses a co-added reference image (\textit{template}) and one recently observed \textit{search image} on which we want to find variability.
The reference image is constructed by stacking a subset of archival images taken in exceptional observing conditions with low sky noise, small PSF size, and high atmosphere transparency (more details in Sec.~\ref{section:dataresults-tmplts}). Since these templates are built from many individual observations, pixels with artifacts (e.g. moving object trails, cosmic rays, CCD blooming and bleeding) are rejected during the co-addition process. In this work, DIA is based on the \cite{alard_method_1998} (herafter A\&L) technique, which uses a kernel to transform the template image such that its pixel locations, orientation, and PSF match the search image.

The \diapipe{} stages are illustrated in the upper panel of Fig.~\ref{fig:pipe_diagram} and briefly described below:
\begin{figure}
   \centering
   \includegraphics[width=.9\hsize]{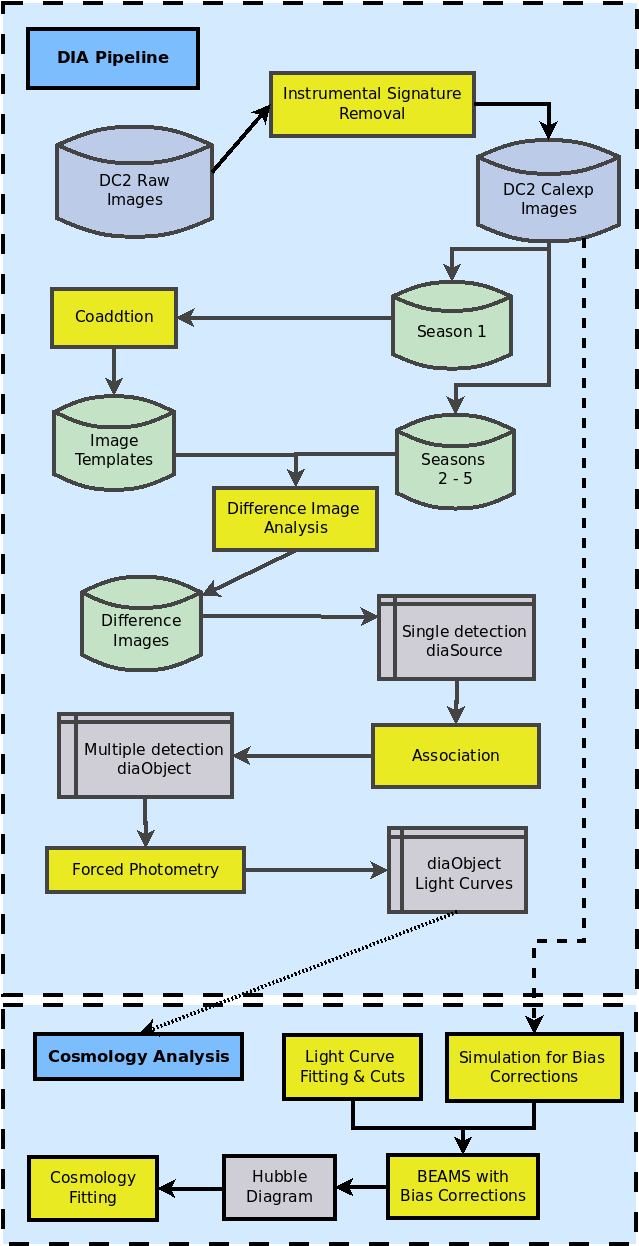}
   \caption{Diagram of \diapipe{} and LSST Science Pipelines (upper box).
   Each processing step (yellow boxes) takes an image (green cylinders) or catalog data (gray tables), and produces new images or catalogs.
   Additionally we show connections to the cosmology analysis (bottom box).
   }
   \label{fig:pipe_diagram}
\end{figure}
\begin{itemize}
   \item \textbf{Instrumental signature removal and calibration}: Simulated images, including calibration frames (dark current, bias and flat field exposures) are ingested to perform Instrumental Signature Removal (or ISR) \add{and image calibration}, resulting in a repository of calibrated image exposures called ``\textit{Calexp's}.''
   \item \textbf{Template Coaddition}: From the Calexp repository, \NimageTemplate{} images are selected from the first season (Y1) with good seeing and low sky noise. Among these images, those that overlap each DC2 patch by a significant fraction are co-added to build templates. \add{Pixel weight was estimated using the inverse variance.}
%
   \item \textbf{DIA}: The A\&L image subtraction algorithm is run on images from seasons 2-5, which produces difference images. \add{Next,} source detection is run on \add{each difference image} to obtain a catalog of DIA single-visit detections, called \diasrc{}. To avoid artifacts near CCD edges, detections within 16~pixels of an image edge are discarded; this cut is about ${\times}3$ larger than the typical PSF-FWHM size. At least one corner of each search image is required to overlap a template image within $65\arcsec$ of the edge.  The DIA kernel basis is composed of 3 Gaussians, with an adaptive spatially varying size to accommodate varying PSF sizes. The basis components also vary spatially.
   \item \textbf{Association}: a candidate-association algorithm creates \diaobjs{} from one or more \texttt{diaSources} that match spatially within $\matchtol$.
   \add{This cut radius is much larger than the average astrometric precision of DC2 calibration (see Fig. 13 of \citep{desc_dc2_2020}).}
   As each \diasrc{} is added to a \diaobj{}, the average RA and Dec coordinate of the \diaobj{} is updated.
   \item \textbf{Forced photometry}: For each \diaobj{}, forced PSF photometry is performed at the location of the object for all overlapping images, regardless of whether there was a \diasrc{} detection. The collection of forced photometry fluxes and uncertainties for each \diaobj{} comprises the light-curve used in the cosmology analysis (Sec.~\ref{section:analysis-cosmology}).
\end{itemize}
%
\add{
\subsection{DIA validation}
\label{section:analysis-diavalidation}
Before using the light curves for the cosmology analysis we perform several validation checks on the performance of the DIA steps explained in Sec.~\ref{section:analysis-pipeframework}.
These validation checks include:
\begin{itemize}
    \item Template quality (depth and PSF size)
    \item Efficiency vs. SNR
    \item Detection depth per band
    \item Artifact contamination level
    \item Photometric precision for fluxes
    \item Photometric uncertainty correlation with Surface Brightness
    \item Photometric flux-outlier fractions
\end{itemize}
}

\subsection{Cosmology analysis}
\label{section:analysis-cosmology}

\add{
Here we describe a cosmology analysis that combines the DC2 light-curves, obtained with the DIA pipeline framework, with a simulated low redshift (\LowZName{}) sample generated with the \snana{} simulation.
The \LowZName\ sample covers a redshift range of $z<0.08$ with an assumed
spectroscopic-selection efficiency of $100\%$.
The ``DC2+\LowZName{}'' analysis}
includes light curve fitting to standardize the SNIa brightness, a Monte Carlo simulation to correct for selection effects, a global fit to produce a bias-corrected Hubble diagram, and a $w$CDM fit to estimate $w$ and $\OM$ (see bottom panel in Fig.~\ref{fig:pipe_diagram}). We closely follow the procedures used in the analyses for Pantheon \citep{Scolnic_pantheon_2018}, PS1 \citep{rest_panstarrs_2014, jones_ps1_2018}, and DES \citep{Brout2019a}.

To the extent possible, we treat DC2 light curve data as real data by not using underlying truth information. However, there are four caveats where truth information is used.
(1) We do not use DC2 data to train the SALT2 model, nor to measure the true color and stretch populations; instead, \add{we use the known SALT2 model for light curve fitting, and we use the known SALT2 and population model for the bias-correction simulations.}
(2) We select light curves from \diaobjs{} that match true DC2 SNe~Ia,  and thus our DC2 sample corresponds to a spectroscopically confirmed sample without contamination from other SNe types.
(3) We use the true DC2 redshifts, and thus assume accurate redshifts from either the SN or correctly-matched host-galaxy.
(4) To characterize DIA detection efficiency vs. \add{Signal-to-Noise Ratio (SNR)} we use the same DC2 light curve data as in the analysis; for future LSST analysis of real data, there will be a separate data stream of fake sources to
measure this DIA property.

The analysis stages described below use programs from the Supernova Analysis (SNANA: Kessler et al. 2009) software package\footnote{\url{https://github.com/RickKessler/SNANA}}:
\begin{itemize}
    \item \textbf{Light curve fitting on data}: we use the SALT2-Extended \citep{pierel_extending_2018} light curve model, the same model used to generate DC2 SNIa, and fit for $t_0$, $x_0$, $x_1$ and $c$ parameters and their covariances. We impose the following selection requirements (cuts) \add{based on previous cosmology analyses}:
        \begin{itemize}
            \item \add{at least one detection (Sec.~\ref{sec:diasources}) in any passband}
            \item maximum $\rm{SNR}>4$ \add{in at least 3 separate passbands}
            \item fitted $|x_1|<3$, and $0<\sigma_{x_1}<2$
            \item fitted $|c|<0.3$
            \item fitted peak MJD uncertainty $<3$ days
            \item fit probability ($P_{\rm{fit}}$), computed from $\chi^2$ and
             the number of degrees of freedom (NDOF), satisfies $P_{\rm{fit}}> 0.05$.
            \item At least one observation before $t_0$, and another 10 days after $t_0$ in the rest frame.
            \item require valid bias correction in BBC (see Hubble Diagram \add{determination} below)
        \end{itemize}
        \add{
    In addition to the SN selection cuts, we select observations that satisfy:
    \begin{itemize}
        \item no interpolated pixels near the center
        \item rest frame time is between $t_0 - 15$ and $t_0 + 45$~days
    \end{itemize}
    } 
    \item \textbf{Simulation for bias corrections}: to prepare for distance bias corrections, we generate a catalog level simulation as described in \cite{Kessler2019}. We use DC2 DIA data to determine a cadence library (Section 6.1 of \citealp{Kessler2019}), detection efficiency vs SNR for each band, and flux uncertainty corrections. The same cadence library and detection efficiency were used to simulate both the DC2 and the \LowZName{} samples.
    \item \textbf{Hubble Diagram}: we use ``BEAMS with Bias Corrections'' (\citealp{Kessler2017}, BBC)  to determine a bias corrected Hubble diagram in \NzBBC{} redshift bins, and to determine nuisance parameters: stretch-luminosity correlation $\alpha$, color-luminosity correlation $\beta$, and intrinsic scatter $\sigma_{\rm{int}}$.
    \item \textbf{Cosmology fitting}: \add{we fit for $w$ and $\OM$ using a fast minimization program that combines the DC2+\LowZName{} SN~Ia Hubble diagram with a cosmic microwave background (CMB) prior using the $\Rcmb$ shift parameter (e.g., see Eq.~69 in \citealp{komatsu_5yrwmap_2009}).}
    \add{To avoid bias from a measured prior, $\Rcmb$ is computed from the DC2 cosmology parameters. To have CMB constraining power similar to that of Planck \citep{ade2016planck}, we tuned the uncertainty $\sigR = \sigRvalue$ as follows. We use the publicly available DES 3-year SN~Ia Hubble  diagram,\footnote{\url{https://des.ncsa.illinois.edu/releases/sn}} which resulted in $\sigma_w({\rm stat}) = 0.042$ when combined with Planck constraints. We run our fast cosmology fitting program on this DES 3-year sample, and tune $\sigR$ to achieve the same $\sigma_w(\rm{stat})$.
    }
\end{itemize}

\section{Data processing results}
\label{section:dataresults}
\begin{figure}
    \centering
    \includegraphics[width=\linewidth]{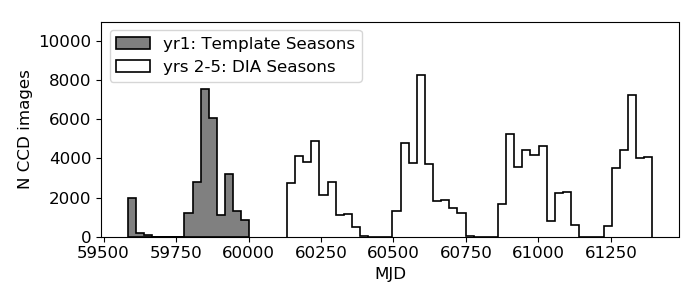}
    \caption{Distribution of observation dates (MJD) from DC2 images spanning 5 seasons. The first season (shaded) is used for templates.}
    \label{fig:seasons}
\end{figure}

We processed DC2 image data using six tracts, which comprises a sky area of \area{} square degrees during 5 observing year-seasons. We used the first year for template creation (Fig.~\ref{fig:seasons}) and the remaining 4 years (hereafter called DIA seasons) for image subtractions, creating a total of ${\sim}106\rm{k}$ DIA images in all LSST filters $ugrizy$.

\subsection{Template creation}
\label{section:dataresults-tmplts}
\begin{figure}
    \includegraphics[width=\linewidth]{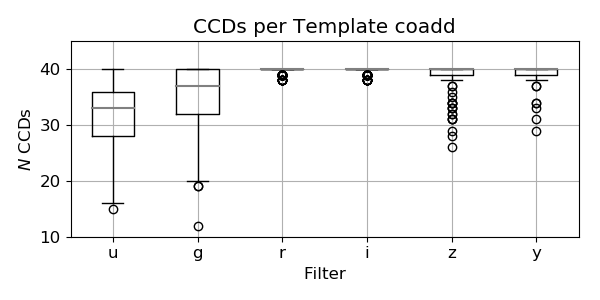}
    \caption{Distribution boxplots of the number of CCD sensors used per template image
    (one template per patch, of roughly 14 arc minutes a side).
    \add{Boxes and whiskers represent $1\sigma$ and $3\sigma$ width of distribution,
    with center line indicating the mean value. Outliers are marked with open circles.}
    }
    \label{fig:ndets_tmpl}
\end{figure}

We show the distributions of the number of images used per template in Fig.~\ref{fig:ndets_tmpl} in a boxplot format. The distributions peak near \NimageTemplate, and they all have a tail extending down to
${\sim}15$ images due to insufficient overlap with the corresponding patch area.

\begin{figure}
    \centering
    \includegraphics[width=\linewidth]{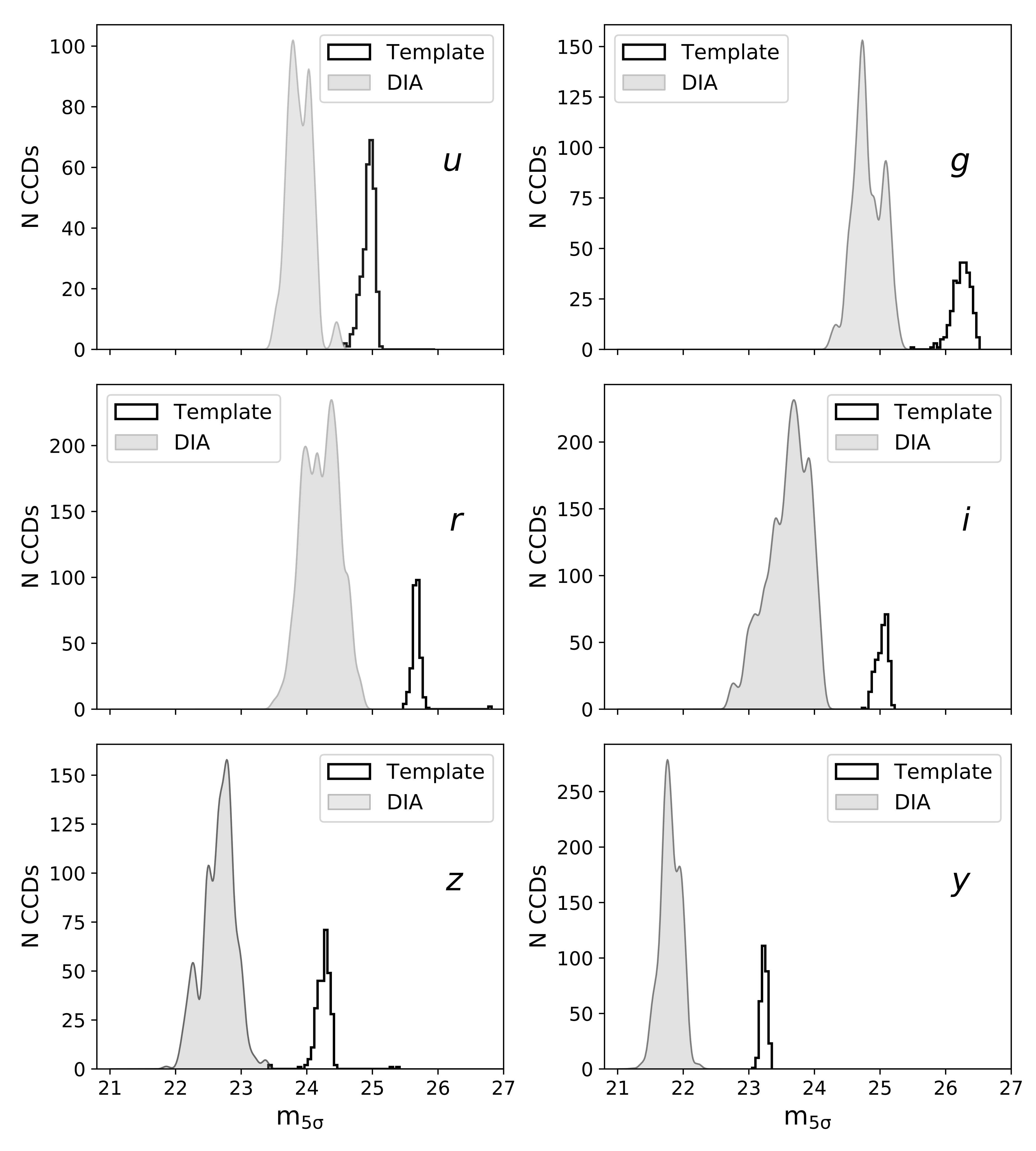}
    \caption{The $\maglim$ distribution for templates, and for visits corresponding to DIA seasons.}
    \label{fig:templatelimitmag}
\end{figure}

\begin{figure}
    \centering
    \includegraphics[width=\linewidth]{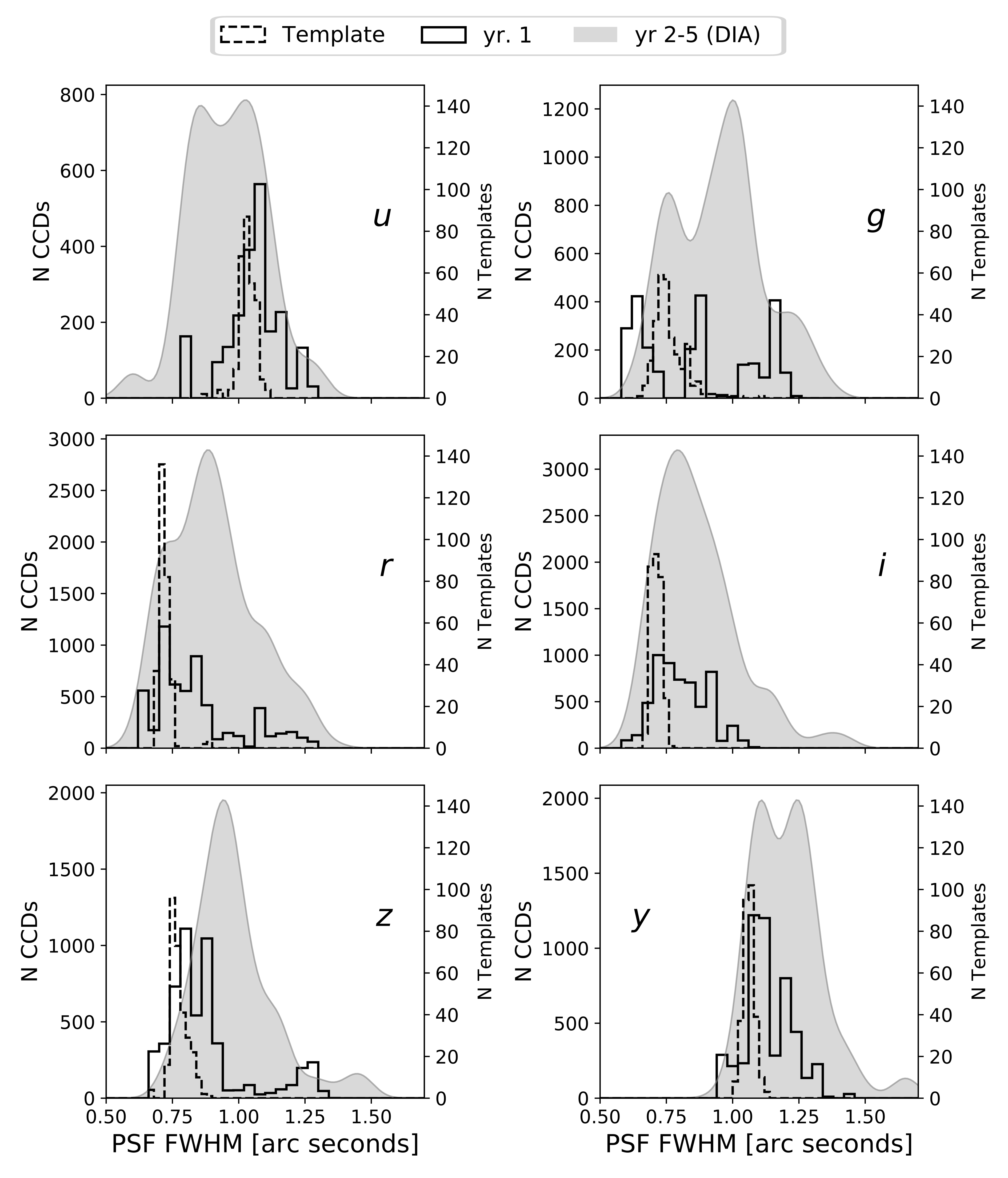}
    \caption{PSF size distribution for \add{template coadds (dashed histogram), year 1 (histogram) and DIA season visits (years 2-5, smooth shaded curve), for each LSST filter.
    Right y-axis correspond to template distribution. For $grizy$, the average template seeing is smaller than for the DIA seasons; for $u$ band the average template seeing is larger.}
    }
    \label{fig:templateseeing}
\end{figure}

For the template and DIA season images, photometric properties of \PSF{} and $5\sigma$ limiting magnitude depth ($\maglim$) are listed in Table~\ref{table:template_table}.
\begin{table*}
    \centering
    \begin{tabular}[t]{@{} l|c|c|*{6}{d{1.2}} @{}}  
    \toprule
    Set & Property & Statistic & $u$ & $g$ & $r$ & $i$ & $z$ & $y$ \\
    \midrule
    Template& $\maglim$ & mean  & 24.94 & 26.24 & 25.69 & 25.03 & 24.25 & 23.23 \\
            &          & RMS   & 0.11 & 0.14 & 0.16 & 0.09 & 0.14 & 0.05 \\
            & \PSF\    & mean  & 1.03 & 0.76 & 0.72 & 0.71 & 0.78 & 1.06 \\
     & [FWHM, arcsec.] & RMS & 0.03 & 0.06 & 0.03 & 0.02 & 0.03 & 0.02 \\
    \midrule
    DIA seasons & $\maglim$ & mean & 23.89 & 24.82 & 24.22 & 23.59 & 22.66 & 21.80 \\
            &              & RMS  & 0.18 & 0.22 & 0.27 & 0.31 & 0.25 & 0.15 \\
            & \PSF\        & mean & 0.96 & 0.95 & 0.90 & 0.87 & 0.98 & 1.21 \\
        & [FWHM, arcsec.]  & RMS & 0.14 & 0.17 & 0.16 & 0.15 & 0.15 & 0.13 \\
    \bottomrule
    \end{tabular}
        \caption{Mean and RMS for properties of images used for template co-addition and for DIA season images.}
        \label{table:template_table}
\end{table*}
We compare $\maglim$ for templates and DIA seasons in Fig.~\ref{fig:templatelimitmag}. The template depth is ${\sim}1$~\magn deeper than DIA season images.
Fig.~\ref{fig:templateseeing} shows that the PSF distribution
for templates is generally smaller and narrower compared to DIA season visits.
The exception is $u$-band, where the template PSF is larger than for DIA season visits because \add{weather fluctuations caused the } first season PSF distribution \add{to be larger} than the other seasons.

\subsection{Image differences}
\label{section:dataresults-diaimages}
\begin{figure*}
   \centering
   \includegraphics[width=\linewidth,trim={0 0 0 2cm},clip]{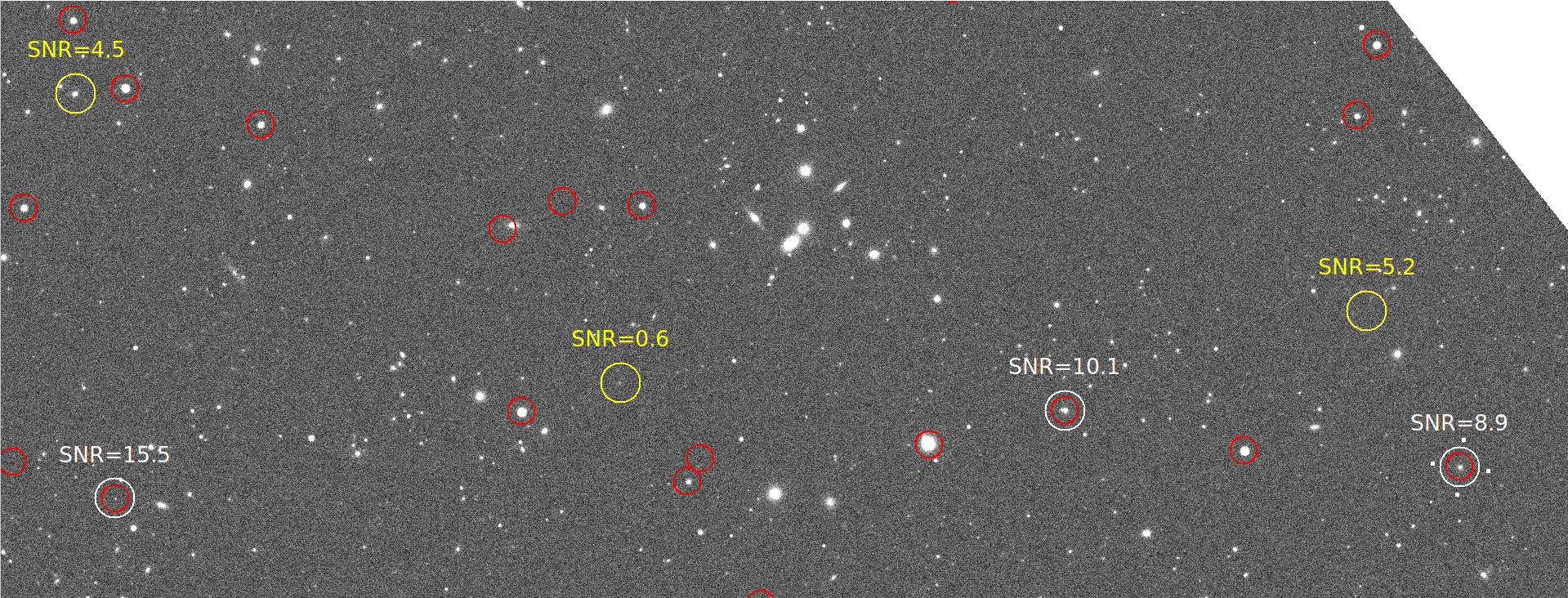}
   \includegraphics[width=\linewidth,trim={0 0 0 2cm},clip]{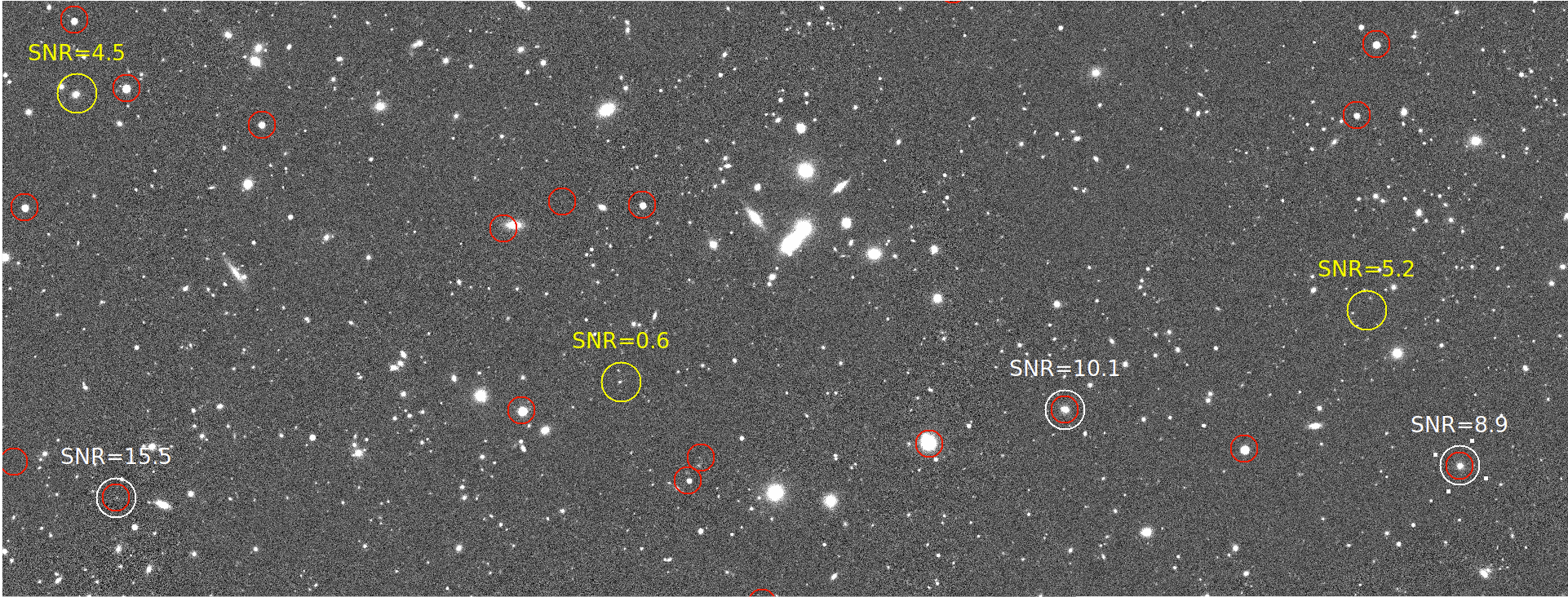}
   \includegraphics[width=\linewidth,trim={0 0 0 2cm},clip]{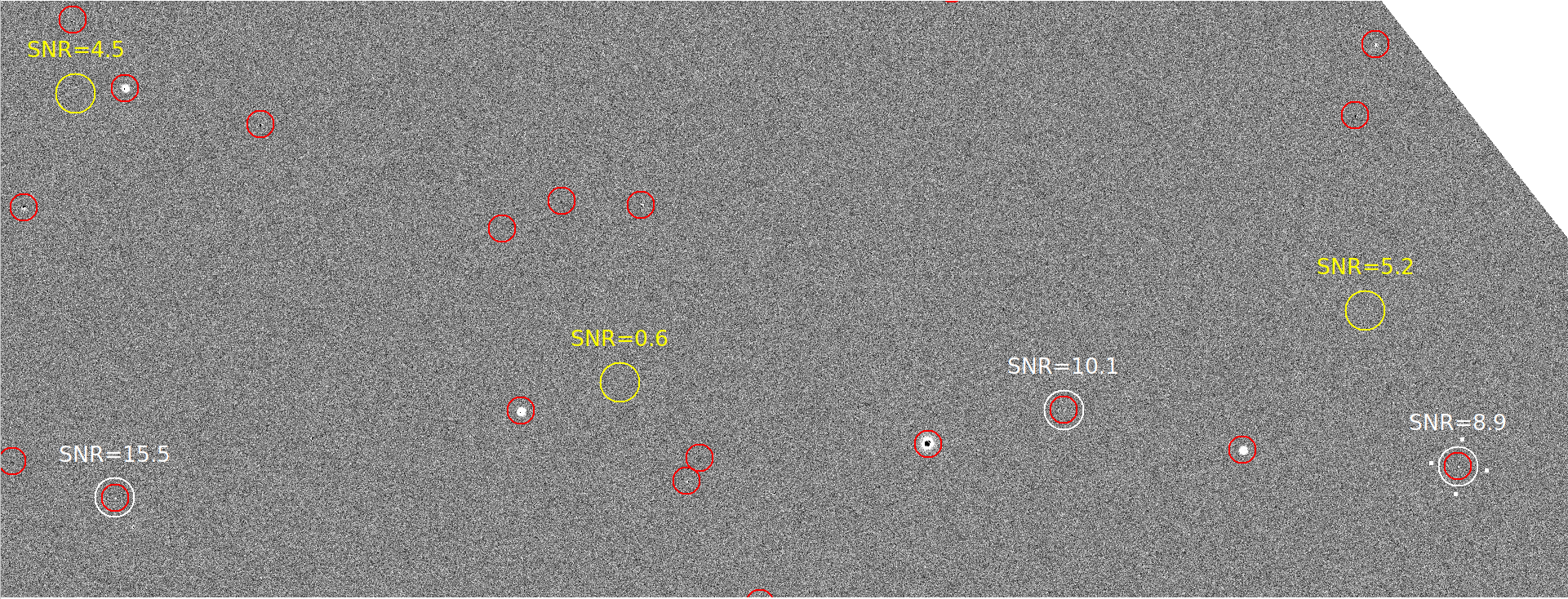}
   \caption{
   \add{
    Example DIA \textit{calexp} image (top), template image (middle), and difference image (bottom) from \diapipe{}.
    Red circles show \diasrc{} detections, white circles show found transients and yellow circles show missed transients.
    On top of the circles we include the SNR value of each true transient point source.
    }}
\label{fig:subs}
\end{figure*}

Our final DIA sample contains a total number of $1967$ visits, or $105,942$ individual images.
In Fig.~\ref{fig:subs}, we show an example of the DIA process from DC2.


\section{Performance metric results}
\label{section:metrics}

Here we characterize the performance of DIA detections, matching, photometric precision, and level of non-astrophysical detections (subtraction artifacts).
For the cosmology analysis, we define metrics based on data-sim distribution overlays of properties of the light-curves, $z$-dependent Hubble diagram bias, and fitted cosmology parameters.

\subsection{DIA on single detections: \diasrcs{}}
\label{sec:diasources}

We crossmatch to the truth catalog in two independent steps. First we use the \diasrc{} catalog as a reference, and find the closest true SN location on the image. Next we use the true SN catalog as the reference, and find the closest \diasrc{}. Finally, we compare these two sets of matches and define a True Positive Match (TP) if there is mutual agreement in both matches, and their separation is below $\matchtol$.
%
\addtwo{If a true SN~Ia doesn't match a \diasrc{} we flag it as a False Negative (FN).}

From this procedure, we find a total of \NmatchSRC{} matches, spanning a wide range of SNR and true magnitudes in all six filters. For the \add{bias-correction} simulation, we measure the detection \eff{} \add{($\effsym=\rm{TP}/(\rm{TP}+\rm{FN})$)} as a function of SNR, and characterize this \eff{} distribution with $\SNRhalf$ defined as $\effsym(\SNRhalf) = 0.5$. To better connect the measured $\effsym$ to simulations, we don't use measured SNR, but instead we compute SNR from the true flux and the true noise, where the latter is computed from the zero point, PSF and sky noise (see Eq 11 in \cite{Kessler2019}).

\begin{figure}
    \centering
    \includegraphics[width=\linewidth]{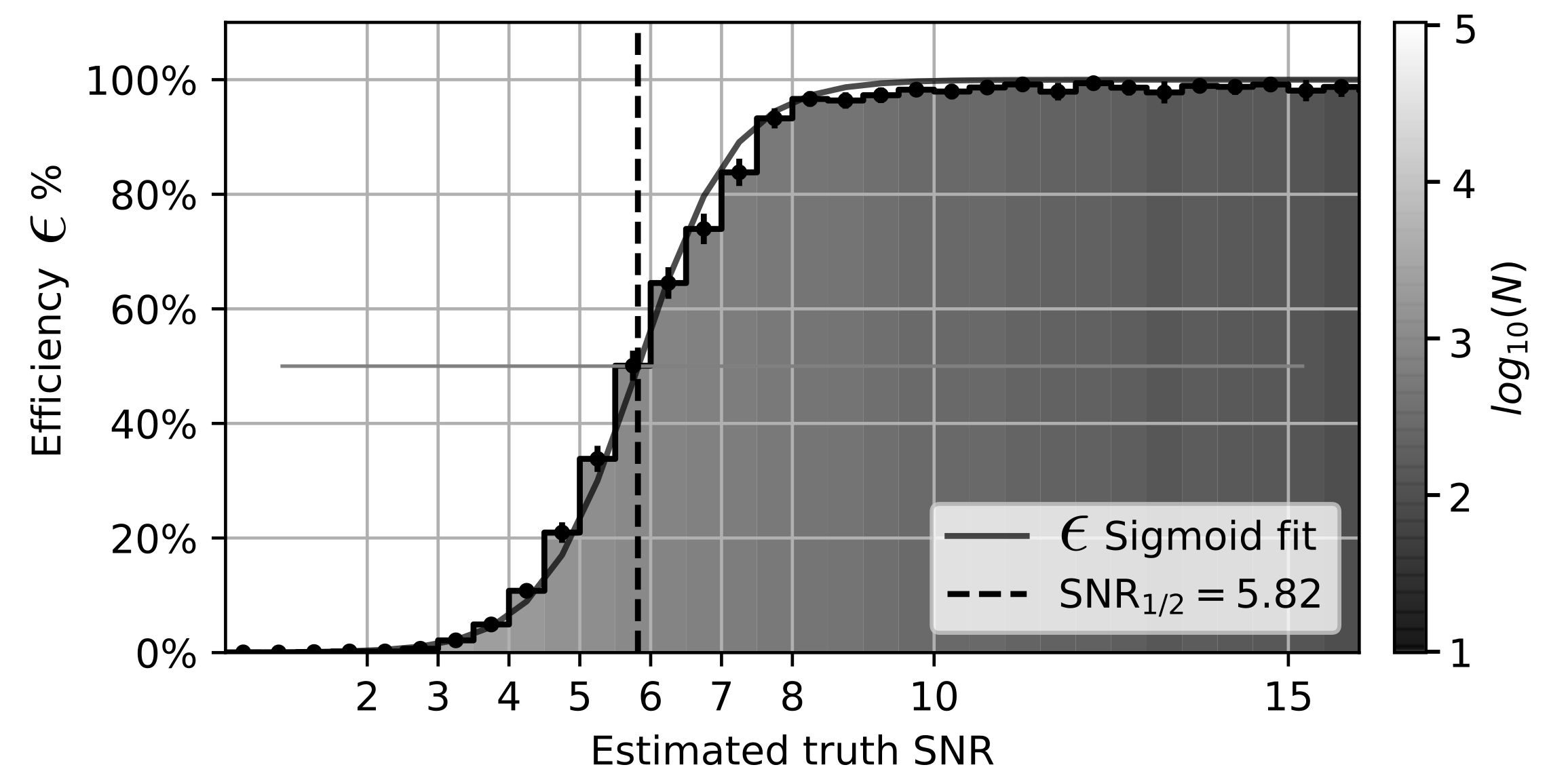}
       \caption{Detection efficiency ($\effsym$)\ vs. \add{calculated} SNR.
       The color scale shows the number of objects per bin, and the dashed vertical line shows $\SNRhalf$.
       }
    \label{fig:eff_vs_SN}
\end{figure}

\begin{figure}
    \centering
    \includegraphics[width=1\linewidth]{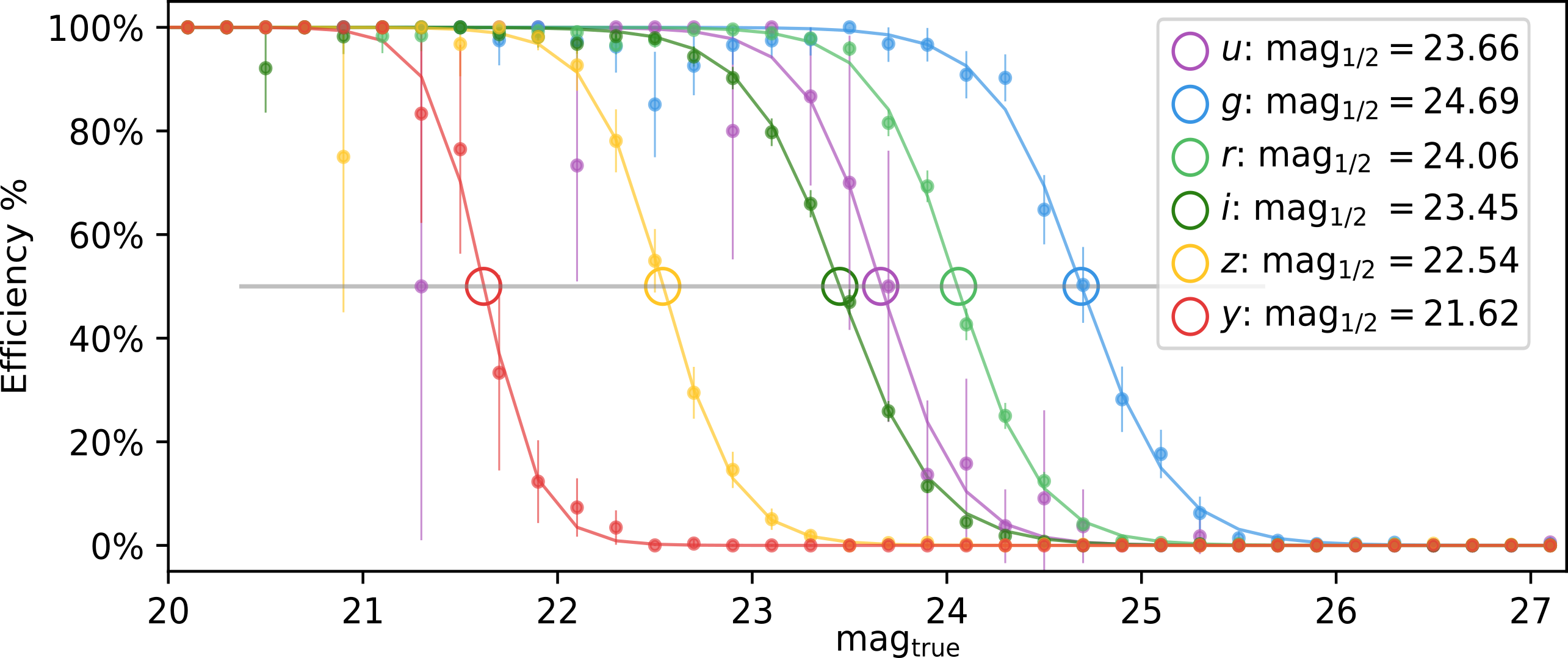}
       \caption{Detection efficiency vs. true magnitude for each bandpass. Open circles indicate $\mhalf$. 
       }
    \label{fig:eff_vs_mag}
\end{figure}

We estimate the value for $\SNRhalf$ by fitting a \textit{sigmoid} function, $\effsym = (1+ e^{-{\rm SNR}})^{-1}$. Fig.~\ref{fig:eff_vs_SN} shows $\effsym({\rm SNR})$ for all filters combined; $\effsym$ increases with SNR as expected, and the filter-averaged $\SNRhalf = 5.82$.
%
\begin{table}
    \centering
    \begin{tabular}{l|cccccc}
    \toprule
    Property & $u$ & $g$ & $r$ & $i$ & $z$ & $y$ \\
    \midrule
    $\SNRhalf$ & 5.76 & 5.57 & 5.87 & 5.84 & 5.59 & 5.60 \\
    $\mhalf$   & 23.66 & 24.69 & 24.06 & 23.45 & 22.54 & 21.62 \\
    \bottomrule
    \end{tabular}
    \caption{Measured $\SNRhalf$ and $\mhalf$ vs. filter.}
    \label{table:eff_table}
\end{table}
We also estimate $\effsym$ as a function of magnitude for each bandpass as shown in Fig.~\ref{fig:eff_vs_mag}, which shows a clear correlation with $5\sigma$ limiting magnitudes (see  Fig.\ref{fig:templatelimitmag}). For each LSST filter, $\SNRhalf$ and $\mhalf$ are listed in Table~\ref{table:eff_table}. The $\SNRhalf$ values are between 5.5 and 6 in each band. The $\mhalf$ values range from 21.6 in $y$ band to $>24$ in the $g$ and $r$ bands.
%

\subsection{Artifact contamination level}
\label{sec:bogus}

To quantify the number of detected artifacts from DIA, we first match \diasrc{} detections to all transients in the DC2 truth catalog, using a tolerance of $1''$.
\add{We define artifacts as unmatched \diasrcs{}.}
Using two bins in PSF seeing size for each filter, Table~\ref{tab:mean_artifx_density} shows the number of \diasrc{} detections, number of matches, and number of artifacts. We show the number of matches for all true \add{astrophysical} variables ($\Nsrcvar{}$) and true SNe~Ia  ($\NsrcSN{}$), along with the percentage of the total number of \diasrcs{} detections. We characterize artifacts by computing the mean and RMS of the density per square degree ($\Bd$).
$\Bd$ is larger for images with smaller PSF, and may in part be due to increased depth for smaller PSF. This effect is most pronounced in $u$ band: $\Bd{\sim}2000$ for PSF$>1\arcsec$, and increases to ${\sim}4000$ for PSF$<1\arcsec$. In $g$ band, $\Bd{\sim}1000$ with a 20\% difference between the PSF bins.
$\Bd$ falls with increasing wavelength, \add{and is correlated with  search depth};
in $y$ band, $\Bd{\sim}300$ and the PSF difference is $<10\%$.
From Table~\ref{tab:mean_artifx_density}, ${\sim}90\%$ of the \diasrc{}
detections are artifacts in the $grizy$ bands; in $u$ band the artifact fraction is 99\% due to template seeing size, which is relatively broader than search image PSF size.
In previous surveys, machine learning methods have significantly reduced artifacts \citep{Goldstein_2015, Kessler2015, Mahabal_MLZTF_2019}, and similar methods are under development within LSST.


\begin{table*}
    \centering
    \caption{Number of \diasrcs{} detections for SN and artifacts, \add{split into two PSF bins}}.
\begin{threeparttable}
\begin{tabular}{c|r|cc|r|r|r|r|rr}
\toprule
Filter & $N_{\rm{CCD}}$\tnote{*} & $\rm{PSF}$ & $\overline{\rm{PSF}}$ & $\Ndiasrc{}$\tnote{+} & $\NsrcSN{}$\tnote{**} & $\Nsrcvar{}$\tnote{$\dagger$}  & $\Nsrcart{}$\tnote{$\ddagger$} & $\overline{\Bd}$ & $\sigma_{\Bd}$ \\
    &       & \multicolumn{2}{c|}{[arcsec]} & & & & &  \multicolumn{2}{c}{[$\rm{deg}^{-2}$]} \\
\midrule
$u$ & 3901  & >1.0 & 1.1 & 1384171 & 61   $(<0.01\%)$ & 6298   $(0.5\%)$  & 1377812 ($99.5\%$) & 2190 & 3092 \\
    & 3902  & <1.0 & 0.9 & 2999820 & 71   $(<0.01\%)$ & 7815   $(0.3\%)$  & 2991934 ($99.7\%$) & 4189 & 6620 \\ \midrule
$g$ & 5870  & >1.0 & 1.1 & 411704  & 692  $(0.17\%)$  & 41423  $(10.1\%)$ & 369589  ($89.8\%$) & 956  & 372  \\
    & 5870  & <1.0 & 0.8 & 502003  & 922  $(0.18\%)$  & 53411  $(10.6\%)$ & 447670  ($89.2\%$) & 1207 & 453  \\ \midrule
$r$ & 13905 & >0.9 & 1.0 & 743402  & 3146 $(0.42\%)$  & 76986  $(10.4\%)$ & 663276  ($89.2\%$) & 725  & 244  \\
    & 13852 & <0.9 & 0.8 & 823973  & 3679 $(0.45\%)$  & 90830  $(11.0\%)$ & 729464  ($88.5\%$) & 841  & 288  \\ \midrule
$i$ & 14346 & >0.8 & 1.0 & 652462  & 2695 $(0.41\%)$  & 70934  $(10.9\%)$ & 578835  ($88.7\%$) & 614  & 227  \\
    & 13495 & <0.8 & 0.7 & 724098  & 3472 $(0.48\%)$  & 76858  $(10.6\%)$ & 643768  ($88.9\%$) & 736  & 250  \\ \midrule
$z$ & 7375  & >1.0 & 1.0 & 279164  & 361  $(0.13\%)$  & 30040  $(10.8\%)$ & 248764  ($89.1\%$) & 497  & 189  \\
    & 6579  & <1.0 & 0.9 & 292950  & 491  $(0.17\%)$  & 31543  $(10.8\%)$ & 260916  ($89.1\%$) & 591  & 227  \\ \midrule
$y$ & 8533  & >1.2 & 1.3 & 177009  & 66   $(0.04\%)$  & 20275  $(11.5\%)$ & 156668  ($88.5\%$) & 293  & 115  \\
    & 8105  & <1.2 & 1.1 & 177733  & 68   $(0.04\%)$  & 22226  $(12.5\%)$ & 155439  ($87.5\%$) & 312  & 122  \\
\bottomrule
\end{tabular}

\begin{tablenotes}\footnotesize
    \item[*] Number of CCD images
    \item[**] Number of \diasrc{} detections
    \item[+] Number of \diasrcs{} matching with SNe (and percentage relative to \diasrcs{})
    \item[$\dagger$] Number of \diasrcs{} matching with other variable sources (and percentage relative to \diasrcs{})
    \item[$\ddagger$] Number of artifacts (and percentage relative to \diasrcs{})
\end{tablenotes}
\end{threeparttable}
\label{tab:mean_artifx_density}
\end{table*}

\subsection{DIA on multiple detections: \diaobjs{}}
\label{sec:diaobjects}
\begin{figure}
    \includegraphics[width=\linewidth]{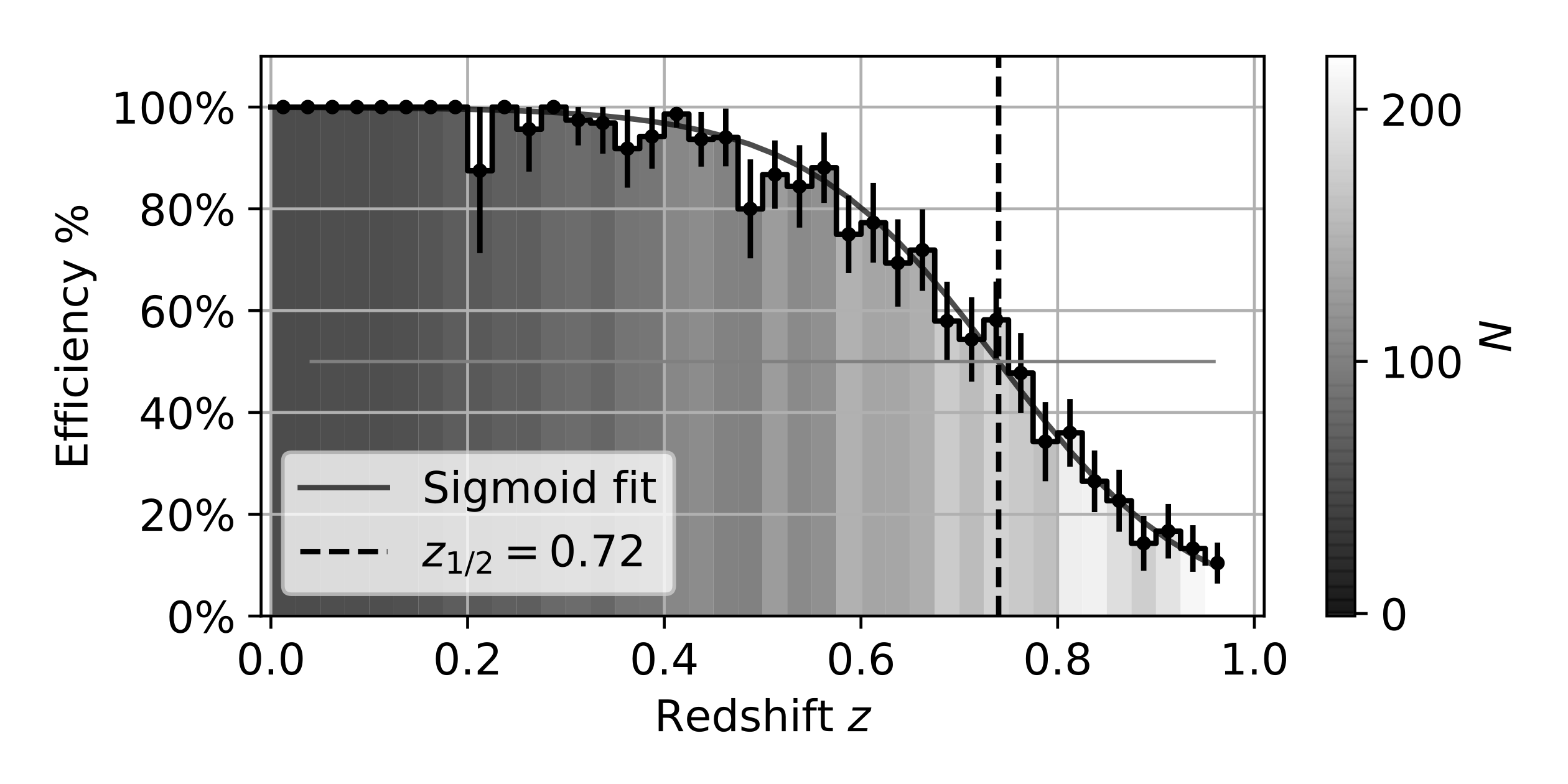}
    \caption{$\effSN$ vs redshift. The dashed vertical line shows where $\effSN=0.5$. Gray color scale reflects the number of events per bin.
    }
    \label{fig:eff_zspace}
\end{figure}

%
We cross-match the \diaobj{} catalog with the true DC2 SNe~Ia using a two step procedure (as explained in Sec.~\ref{sec:diasources}) with a tolerance radius of $\matchtol$, finding a total of $\NmatchSN$ matched SNe. Fig.~\ref{fig:sky_area} shows the \SKYNAME{} area, and true SNe that were matched and not matched to a \diaobj{}. We define SN detection efficiency ($\effSN$) as the probability of associating a true SNIa with a \diaobj{}. Fig.~\ref{fig:eff_zspace} shows $\effSN$ vs. redshift for a subsample of true SNe~Ia which have been observed more than 5 times, have at least one observation before $t_0$, and have at least one observation after $t_0+10$ days in the rest-frame. Fitting this distribution to a sigmoid model (as in Sec.~\ref{sec:diasources}), $\effSN=0.5$ at $z=0.72$.
%
\subsection{DIA photometry: flux measurements}
\label{sec:diafluxes}

\begin{figure}
    \includegraphics[width=\linewidth]{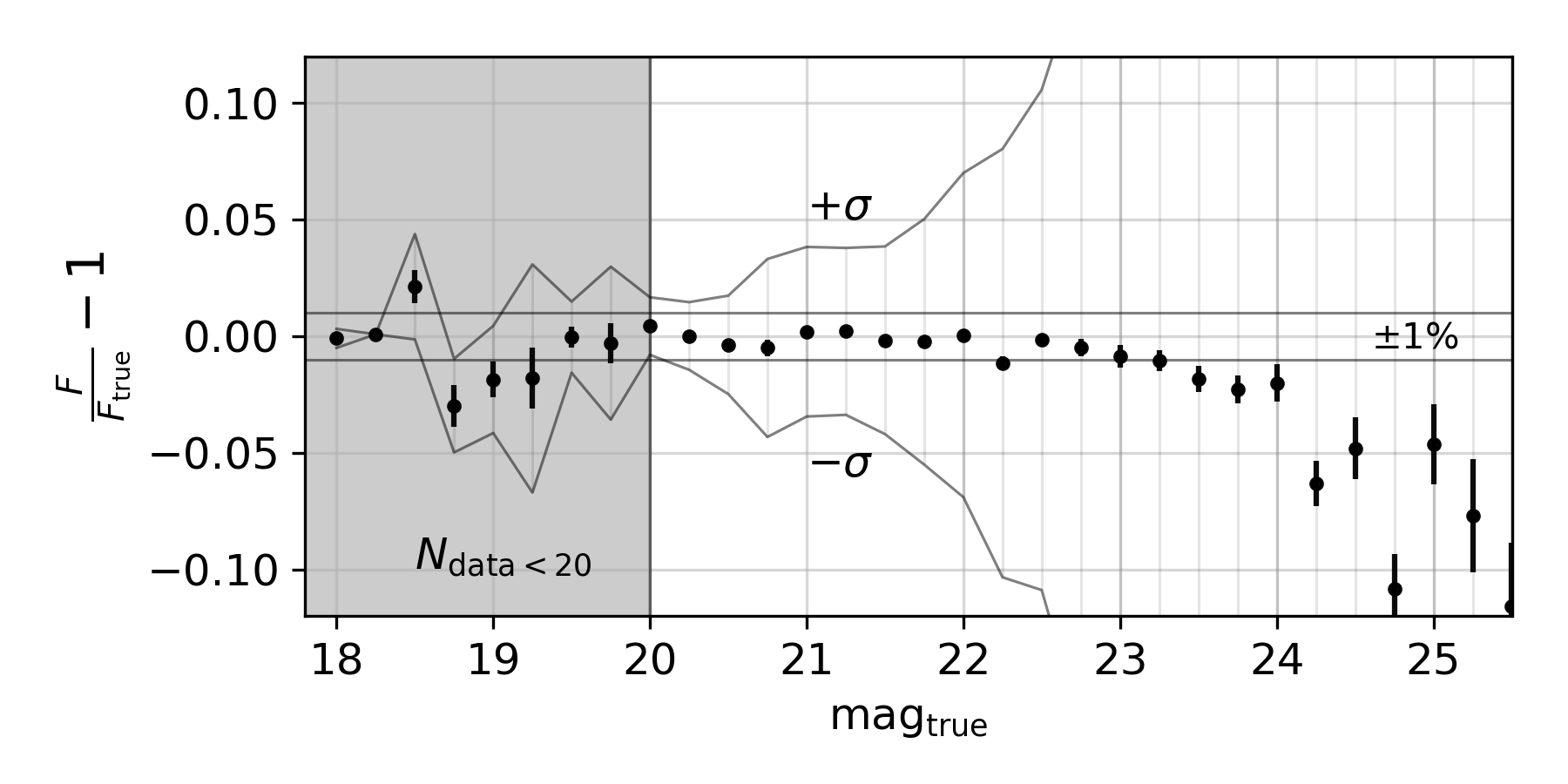}
    \caption{Mean fractional photometric bias, $F/\Ftrue-1$, as a function of true SN magnitude. Error bars show uncertainty on the mean, and solid gray lines show the standard deviation ($\pm 1\sigma$) in each bin, as well as the $\pm1\%$ margin. The shaded area shows bins with less than 20 epochs per bin
    (a total of 75 observations); these epochs are all from 7 SNe.
    }
    \label{fig:mag_vs_mag_resid}
\end{figure}

Forced PSF photometry is measured at the \diaobj{} location on all DIA images. Using the set of \diaobjs{} matched with DC2 SNe, we measure flux and magnitude residuals. Fig.~\ref{fig:mag_vs_mag_resid} shows the fractional photometric bias as a function of true SN magnitude \add{($\mtrue$)},
and the RMS in each bin is illustrated by the $\pm 1\sigma$ envelope.
The shaded region shows low-statistics bins with 20 observations, but only 7 events.
While there is a hint of bias for bright events, note that correlated residuals among observations
from the same event would result in under-estimated uncertainties.
We accurately measure fluxes \add{for $19 < \mtrue < 23.25$}, where the mean fractional photometric bias values \add{are} $<1\%$. For magnitudes \add{$\mtrue > 23.25$} the photometry is biased towards faint values, suggesting a slight bias in the sky subtraction.

\subsection{DIA photometry: flux uncertainties}
\label{sec:diafluxerr}
\begin{figure}
    \centering
    \includegraphics[width=\linewidth]{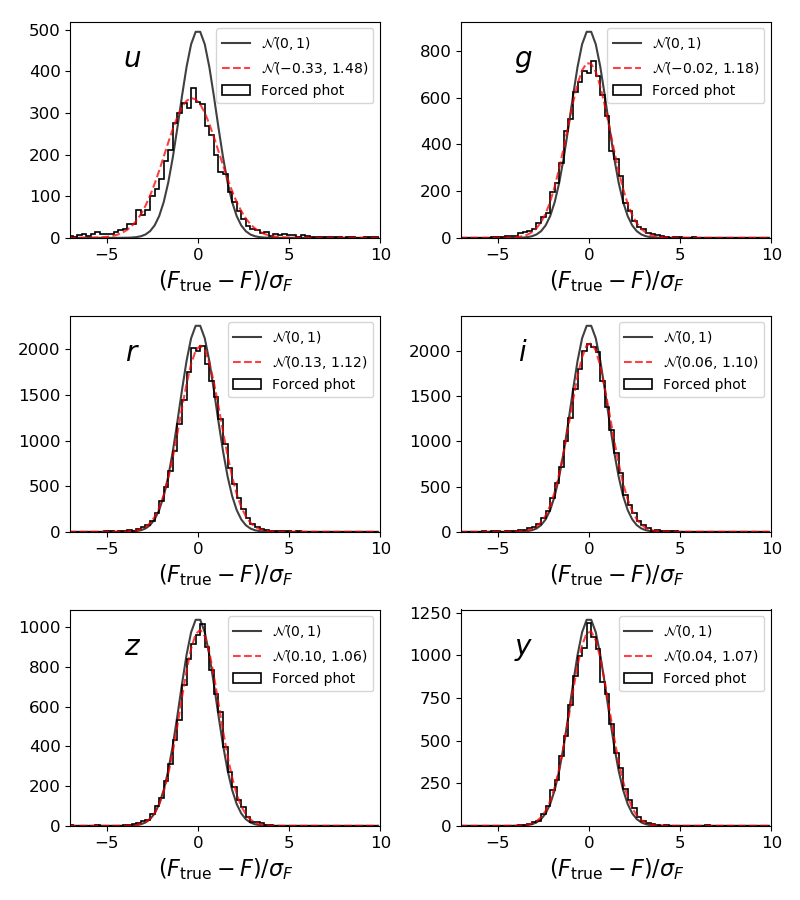}
        \caption{Distribution of Forced photometry pull values as described in the text.
        Red curve shows best fit Gaussian model,
        and black curve is a zero-mean unit-dispersion Normal distribution for comparison.
        }
    \label{fig:flux_pulls_perband}
\end{figure}

To evaluate the flux uncertainties, we measure the pull distribution in each band (Fig.\ref{fig:flux_pulls_perband}), $(F-\Ftrue)/\sigF$, where $F$ is the forced photometry flux, $\sigF$ is the uncertainty, and $\Ftrue$ is the true flux.
Defining
\begin{equation}
  \RMSpull \equiv {\rm RMS}[(\Ftrue-F)/\sigF]~,
  \label{eq:RMSpull}
\end{equation}
we expect $\RMSpull = 1$ if the uncertainties are accurate. We find that the distributions are nearly Gaussian, but $\RMSpull > 1$. For $u$-band ${\RMSpull \sim}1.5$, indicating a significant underestimate of the flux uncertainties. For the other bands, $\RMSpull {\sim}1.1$.

\begin{figure}
    \centering
    \includegraphics[width=\linewidth]{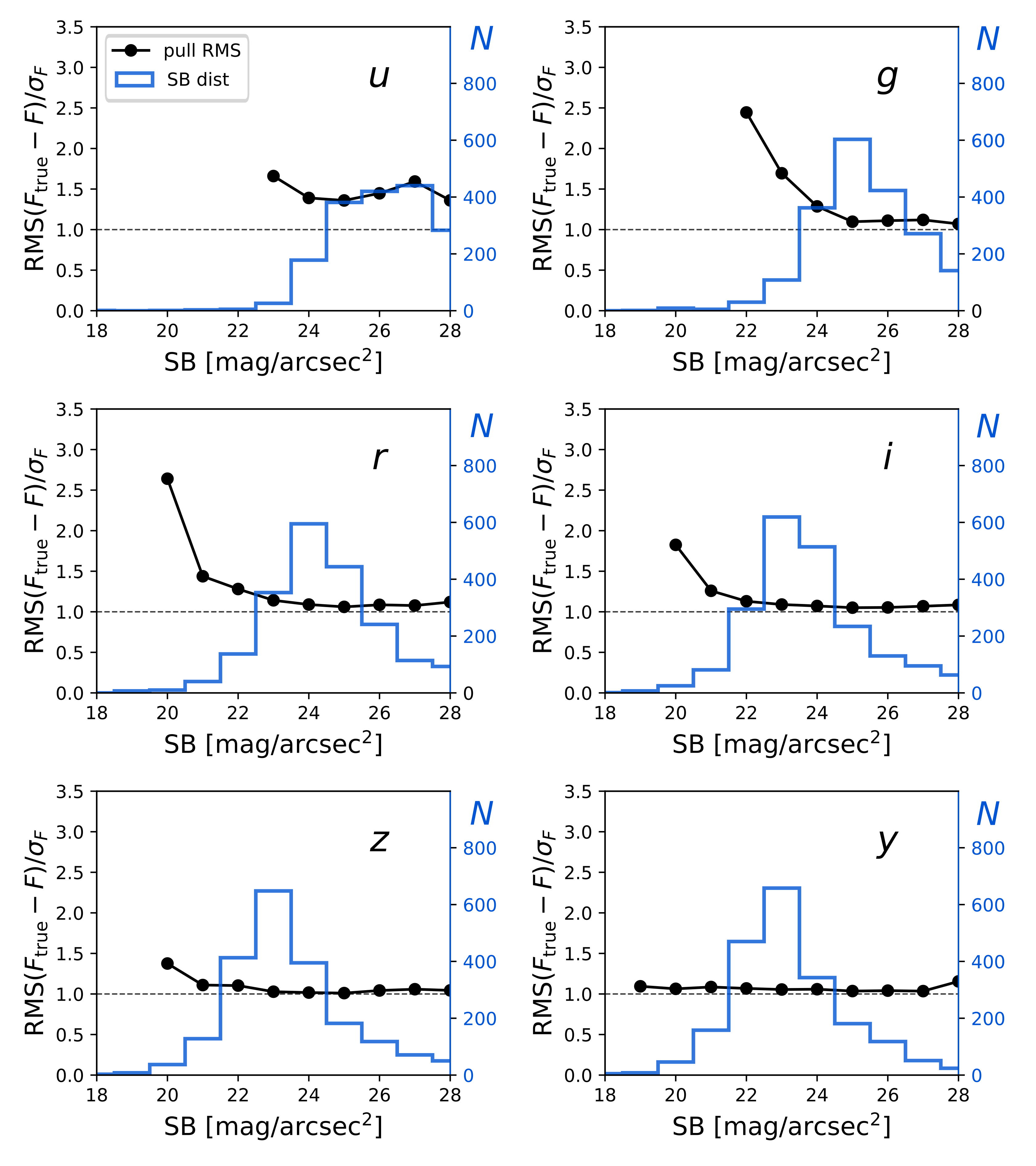}
    \caption{
        RMS of forced photometry pull as a function of \add{$\mSB$ at the location of each SN.}
        Dashed line shows the unit dispersion reference.
        \add{The distribution of $\mSB$ is also shown for each band.}
    }
\label{fig:flux_pull_rms_vs_sbmag}
\end{figure}

For DES, \citet{Kessler2015} reported a ``Surface Brightness (SB) anomaly'' in which the true scatter was larger than the reported uncertainties, and this effect was strongly correlated with SB \add{at the SN location}.
Here we repeat this analysis for DC2 and measure the local surface brightness magnitude ($\mSB$) in template coadds at each SN location, using aperture photometry with a radius of $0''.9$.
\add{We find that the SB anomaly is present in} DC2 simulated images.
Fig.~\ref{fig:flux_pull_rms_vs_sbmag} shows $\RMSpull$ vs $\mSB$ for each filter. $\RMSpull$ is near 1 for faint SB, and increases with increasing SB for $grizy$ bands. For $u$ band, $\RMSpull  \sim 1.5$ for all $\mSB$. $\RMSpull$ reaches a maximum of $\sim 3$ in the $g$ and $r$ bands with $\mSB \sim 21$ and $\mSB \sim 20$~\magn respectively. $\RMSpull$ values are consistent for SNe of all peak brightness.
\add{This effect is not understood and we therefore apply an empirical scale correction (Fig.~\ref{fig:flux_pull_rms_vs_sbmag}) to the flux uncertainties.
}
\add{Comparisons with DES are presented in Sec.\ref{section:discussion}.}

We show in Table~\ref{tab:outliers} the robust mean and $\RMSpull$ for each filter flux pull distribution, as well as the percent of $5\sigma$ and $10\sigma$ outliers. The $u$ band outlier fraction (${\sim}1\%$) is roughly an order of magnitude higher than in the other bands.
\begin{table}
    \centering
    \caption{Flux pull distribution parameters}
    \begin{threeparttable}
    \begin{tabular}{l|ccccc}
        \toprule
        Filter & N & $\overline{x}$\tnote{a} & $\RMSpull$ & $f_{5\sigma}$ (\%)\tnote{b}  & $f_{10\sigma}$ (\%)\tnote{c}  \\
        \midrule
        $u$ & 3936 & -0.35 & 1.48 & 4.9 & 2.0 \\
        $g$ & 8480 & -0.03 & 1.18 & 0.24 & 0.09 \\
        $r$ & 22366 & 0.15 & 1.12 & 0.22 & 0.05 \\
        $i$ & 22530 & 0.06 & 1.10 & 0.16 & 0.04 \\
        $z$ & 10221 & 0.10 & 1.06 & 0.06 & 0.01 \\
        $y$ & 11892 & 0.05 & 1.07 & 0.07 & 0.02 \\
        \bottomrule
    \end{tabular}
\begin{tablenotes}\footnotesize
    \item[a]{Robust mean of the pull}
    \item[b]{Percent of $5\sigma$ outliers.}
    \item[c]{Percent of $10\sigma$ outliers.}
\end{tablenotes}
\end{threeparttable}
    \label{tab:outliers}
\end{table}
We found $u$ band bias and $\RMSpull$ to be anti-correlated with PSF size; the smallest PSF bin has the largest bias and $\RMSpull$.
%


\subsection{Cosmology Analysis Results}
\label{section:sn1acosmology}
Following the steps for cosmology analysis described in Sec~\ref{section:analysis-cosmology} and shown visually in Fig.~\ref{fig:pipe_diagram}, we apply the selection requirements and fit light curves with the SALT2 light curve model; a total of \NDATACUTS{} events pass cuts (\NLSSTCUTS{} for DC2, and \NLOWZCUTS{} for \LowZName{}).
\add{We also create two catalog-level simulations that have the same DC2/\LowZName{} proportion as the data, }and undergo the same cuts and light curve fitting as the data:
(1) a \nameSimData{} simulation with \NSIMCUTS{} events (\NSIMDATACUTS{} for DC2 and \NSIMLOWZCUTS{} for \LowZName{}) is used to compare data-sim distributions and to crosscheck the analysis, and
(2) a large ($\NBIASCOR{}$ events) bias-correction simulation is used in BBC.
The difference between the two simulations is that the latter is generated on a $2\times 2$ grid of $\alpha$ and $\beta$ to enable interpolating the bias-correction during the BBC fit.

\begin{figure}
    \centering
    \includegraphics[width=0.95\linewidth]{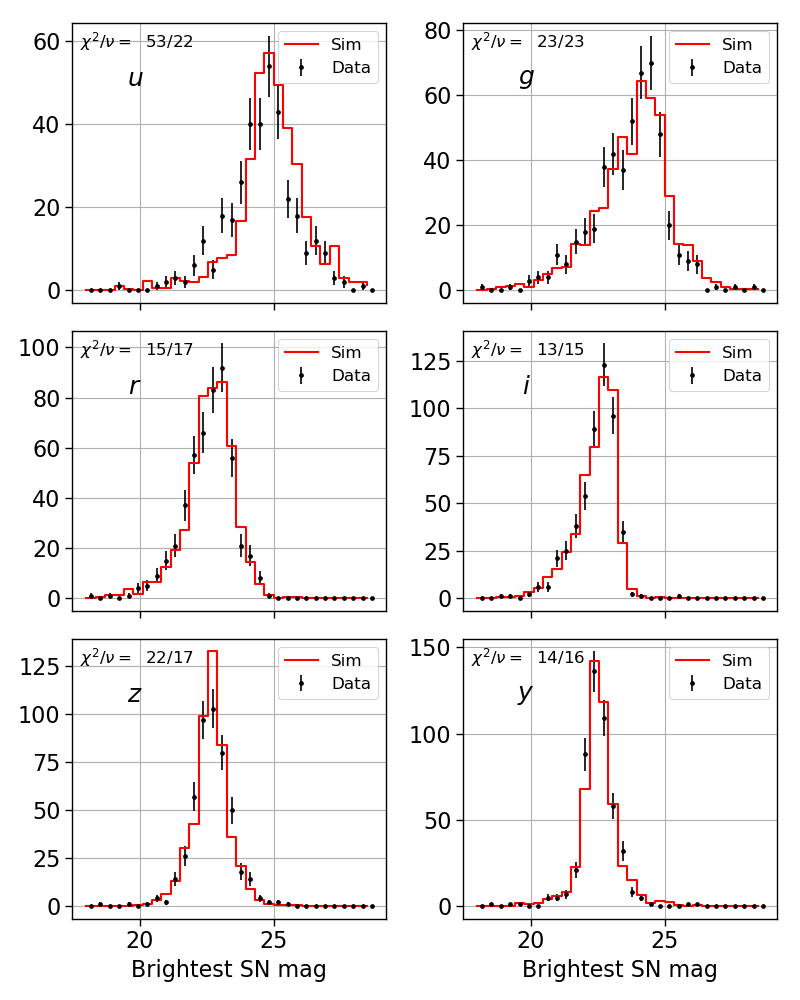}
    \caption{
    Distribution of brightest observed \magn in each filter for DC2 data (black circles) and \nameSimData{} simulation (red histogram).  Each simulated distribution is scaled to match the DC2 sample size. The $\chi^2$ per degrees of freedom ($\nu$) quantifies the data-sim agreement.
    }
    \label{fig:peak_brightness}
 \end{figure}

\begin{figure}
    \includegraphics[width=\linewidth]{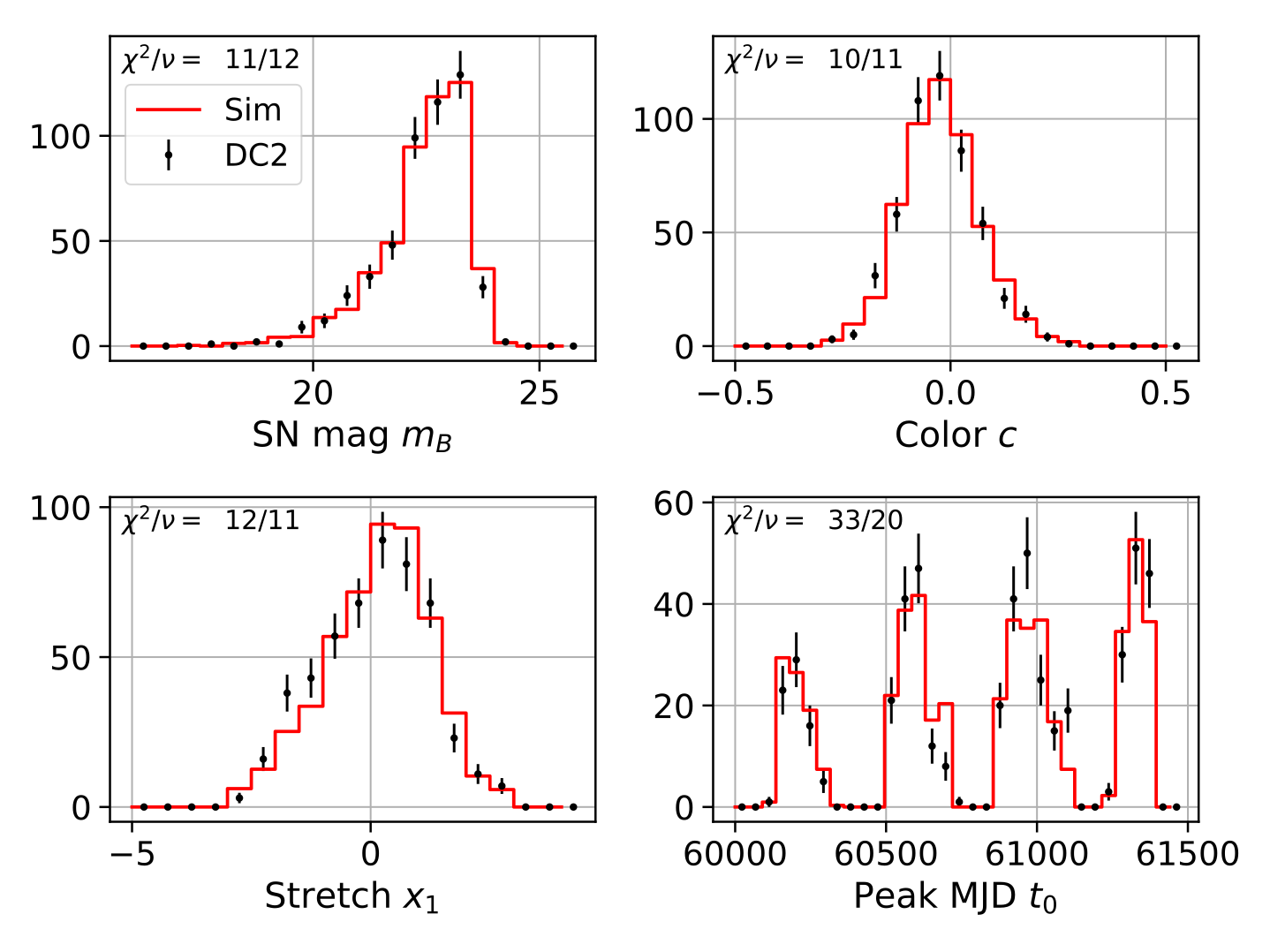}
    \caption{
     Distribution of SALT2 fit parameters for DC2 data (black circles) and \nameSimData{} simulation (red histogram). Each simulated distribution is scaled to match the DC2 sample size. The $\chi^2$ per degrees of freedom ($\nu$) quantifies the data-sim agreement.
    }
    \label{fig:fitted_dc2sim}
 \end{figure}

\add{
Since the \LowZName{} sample is generated by the same catalog simulation used for bias-correction,
there is no need to validate this bias-correction. However it is important to validate the
bias-correction for DC2 by comparing several distributions}
between the DC2 data and the \nameSimData{} simulation:
the observed magnitude for the brightest flux in each filter (Fig.~\ref{fig:peak_brightness}),
and SALT2 fit parameters (Fig.~\ref{fig:fitted_dc2sim}).
\add{All distributions show excellent agreement
except for brightest $u$-band mag in Fig.~\ref{fig:peak_brightness}.}

\begin{figure}[hb]
    \includegraphics[width=0.9\linewidth]{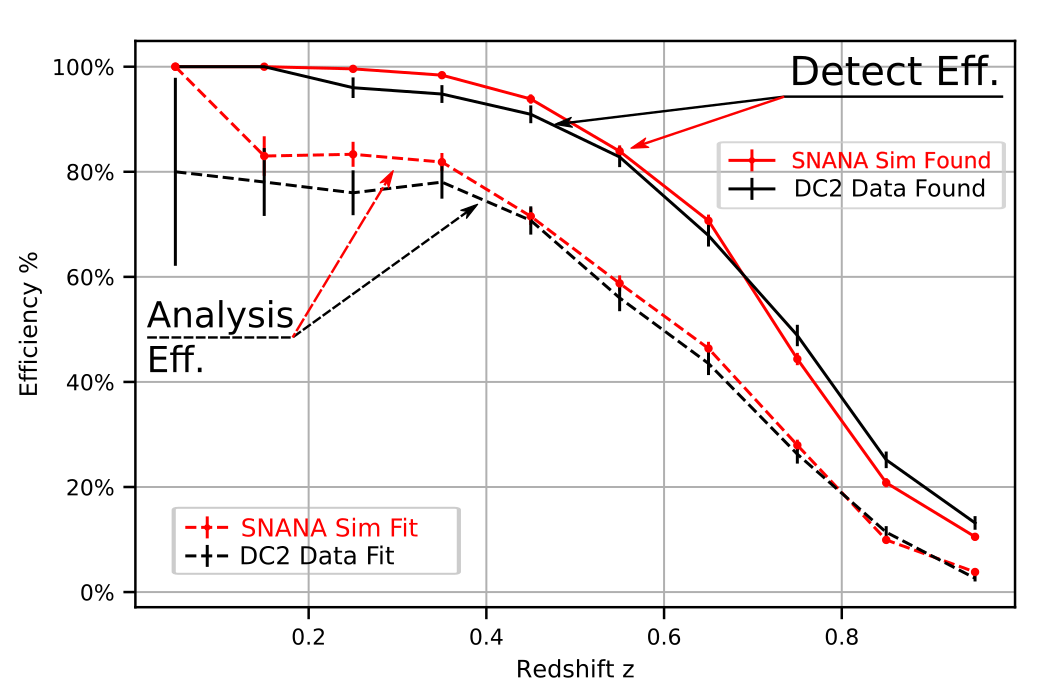}
    \caption{
    Detection efficiency vs. redshift in DC2 data (black solid curve), and in \nameSimData{} simulation (red solid). Analysis efficiency vs. redshift for events satisfying cuts and SALT2 fits in DC2 data (black dashed curve) and in \nameSimData{} simulation (red dashed curve).
    }
    \label{fig:eff_fitting}
\end{figure}

In Fig.~\ref{fig:eff_fitting} we compare the DC2 detection efficiency vs. redshift, as well as the analysis efficiency vs redshift using the requirements listed in Sec.~\ref{section:analysis-cosmology}.
\add{
Fitting a sigmoid function to the detection efficiency curves,
$\zhalfsym = \zhalfDetect\pm 0.0051$ for DC2 data,  and
$\zhalfsym = \zhalfDetectSim\pm 0.003$ for the \nameSimData{} simulation. }
%
For the analysis efficiency, $\zhalfsym = \zhalfCuts\pm0.02$ for
DC2 data, and $\zhalfsym = \zhalfCutsSim\pm0.01$ for the \nameSimData{} simulation.
%

\begin{figure}
    \includegraphics[width=\linewidth]{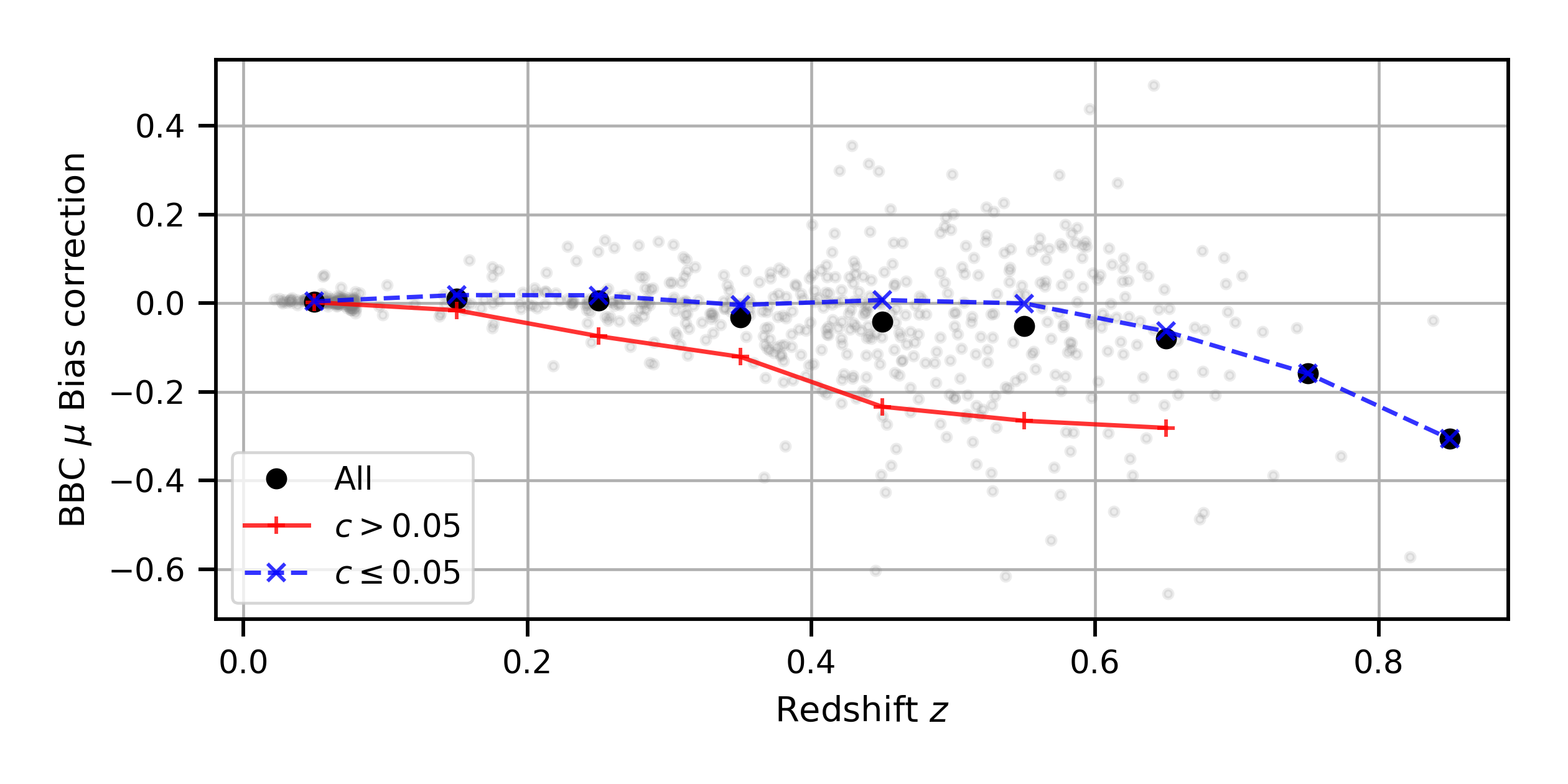}
    \caption{ \add{Average DC2} bias correction \add{vs. redshift for BBC fitted distances.}
    \add{Black dots are for all events, blue curve is for fitted SALT2 color $c<0.05$,
    red curve is for $c>0.05$.}
    }
    \label{fig:bbc_biascor}
\end{figure}

\add{The distance bias correction for \LowZName{} averages to zero,
with small ${\sim}0.01$~mag fluctuations.}
For DC2, the distance bias correction vs. redshift is shown in Fig.~\ref{fig:bbc_biascor} for all events (black circles), where the average bias \add{increases rapidly for $z > 0.6$.}
The subset of blue ($c<0.05$) events, which are brighter than average,
\add{has a smaller bias in the intermediate redshift range.}
The fainter subset of red ($c \leq 0.05$) events has a much larger \add{bias at lower redshifts.}
Accurate simulations and bias corrections are essential for the cosmology analysis.

\begin{figure}
    \includegraphics[width=\linewidth]{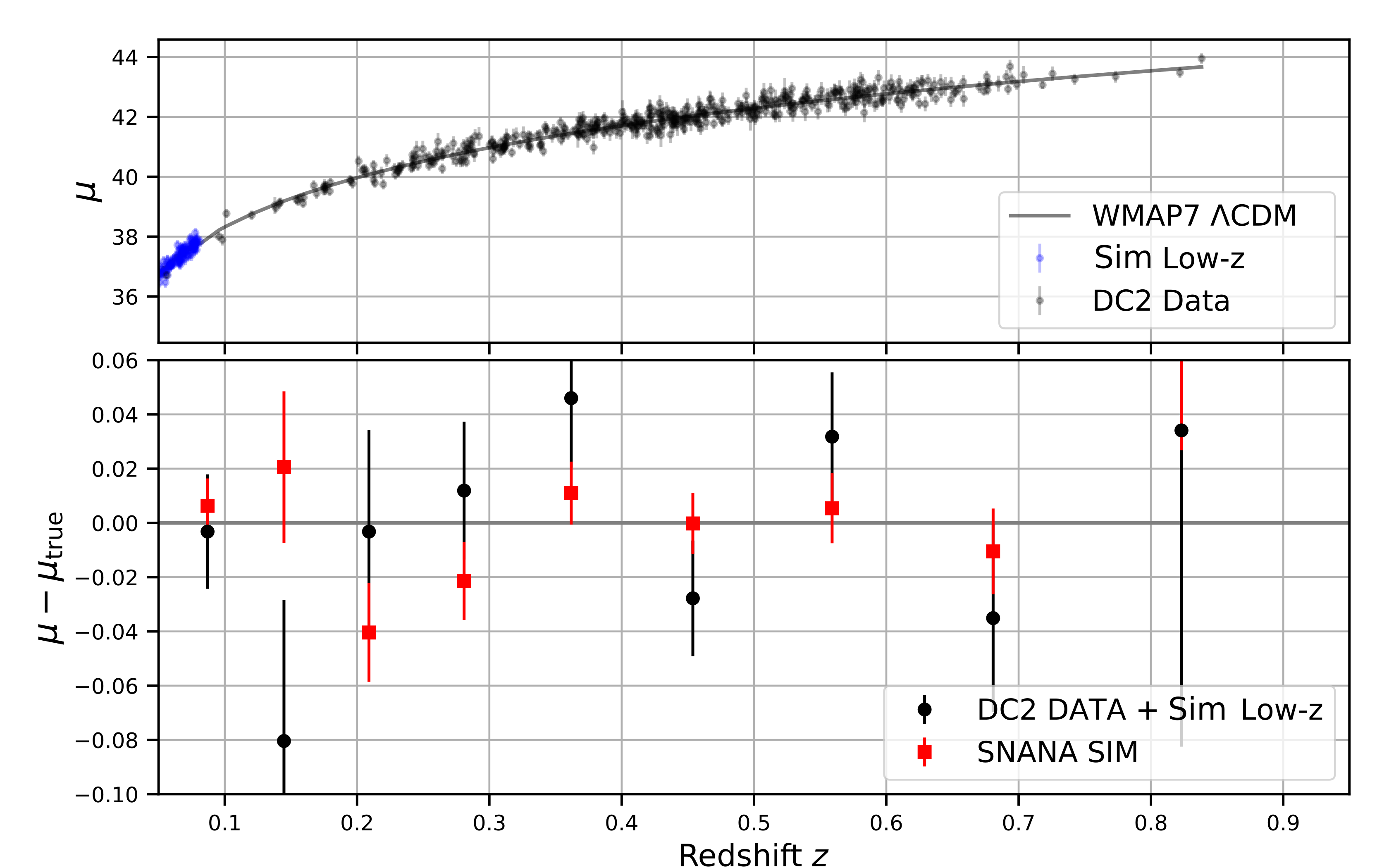}
    \caption{Bias corrected Hubble Diagram for DC2 data (top) and redshift-binned Hubble residuals for DC2 data and DC2-like simulation (bottom).}
    \label{fig:hubble_diagram_hresiduals}
\end{figure}

\add{For the full DC2+\LowZName{} sample,} Fig.~\ref{fig:hubble_diagram_hresiduals} shows the bias corrected Hubble diagram, and the Hubble residuals with respect to the reference DC2 cosmology, for both DC2+\LowZName{} data and \nameSimData{} simulation. Using \NzBBC{} bins of redshift we estimate mean Hubble residuals and error on the mean, shown in the lower panels of Fig.~\ref{fig:hubble_diagram_hresiduals}; the binned residuals are $<0.04$~\magn, and consistent with 0, and no clear trend is seen with redshift.

\newcommand{\wbiasDATA}{-0.032 \pm 0.046}
\newcommand{\ombiasDATA}{-0.007 \pm 0.013}
\newcommand{\wbiasSIM}{-0.002 \pm 0.026}
\newcommand{\ombiasSIM}{0.001 \pm 0.009}

After applying the bias correction analysis to the \nameSimData{} simulation,
we measure nuisance and cosmological parameters.
\add{Defining the bias on $x$ as $\Delta x \equiv x - x_{\rm true}$,}
the biases with respect to input values (Tab.~\ref{tab:salt2dc2pars}) are shown in
\add{the ``\nameSimData{} Sim'' row of}   Table~\ref{tab:BBCparameter_estimation},
and these biases are consistent with zero:
\add{$\Delta w = \wbiasSIM$ and $\Delta\OM = \ombiasSIM$.}
After this validation, we apply the same treatment to DC2+\LowZName{} data and obtain
nuisance and cosmological parameters (\add{DC2 row in} Table~\ref{tab:BBCparameter_estimation}). We find $\Delta w = \wbiasDATA$ and $\Delta\OM = \ombiasDATA$. \add{The nuisance and cosmological parameter biases are consistent with zero.}

\begin{table*}[h]
    \centering
    \caption{Bias corrected parameter estimation for DC2 sample and simulations.}
    \begin{tabular}{l|c|ccc|ccc}
    \toprule
     & & \multicolumn{3}{c|}{From Cosmology Fit} &  \multicolumn{3}{c}{From BBC Fit} \\
    Data Set & $N_{\rm{Events}}$ & $\Delta w=w-w_{\rm{true}}$  & $\Delta\OM=\OM-\OM^{\rm{true}}$  & $\chi^2/\nu$ & $\Delta\alpha = \alpha-\alpha_{\rm{true}}$ & $\Delta\beta = \beta-\beta_{\rm{true}}$ &
    $\sigint / \sigint^{\rm{true}}$ \\ \midrule
    DC2+\LowZName{} & \NDATACUTS{} & $\wbiasDATA$ & $\ombiasDATA$ & 11/8 & $-0.004 \pm 0.010$ & $-0.15 \pm 0.14$ & $1.02$  \\ \midrule
    \add{\nameSimData{}} Sim & \NSIMCUTS{} & $\wbiasSIM$ & $\ombiasSIM$ & 12/8 & $0.004 \pm 0.005$ & $-0.12 \pm 0.06$ & $0.94$  \\ \bottomrule
    \end{tabular}
    \label{tab:BBCparameter_estimation}
\end{table*}

\section{Discussion}
\label{section:discussion}

Our DIA performance study is similar to that in \citet{Kessler2015} for the Dark Energy Survey (DES),
where they used fake light curves injected onto real images during DES operations.
\add{In Table~\ref{tab:DC2_vs_DES} we compare several difference-image properties for
DC2 and DES. To avoid confusion related to different search depths,
we only compare DES bands with similar DC2 depth in Table~\ref{table:eff_table}:
DES $g$-band in their deep fields ($\mhalf{\sim}24.5$~mag)
and DES $i$-band in their shallow fields ($\mhalf{\sim}23.5$~mag).}

\add{
For SNR at 50\% detection efficiency,
the $\SNRhalf$ values are almost identical in $g$-band,
while the DES $i$-band value is ${\sim}10$\% smaller.
To compare the excess scatter on bright galaxies, we evaluate the SB magnitude
where the flux-uncertainty scale factor is 2: $\mSBcompare$.
These $\mSB$ values are nearly identical in $g$-band
(22.1 and 22.0~mag for DC2 and DES, respectively),
but differ by 0.7~mag in the $i$-band (19.8 and 21.5~mag).
The $\mSBcompare$ values for all DC2 bands are shown in
Table~\ref{tab:SBRMS}.
The number of detection artifacts per deg$^2$ is also similar;
DC2 is a factor of 2 higher for $g$-band, and 7\% smaller in $i$-band.
Finally, the $5\sigma$ flux-outlier fractions for DC2 are about a factor of 2
smaller compared to DES; the corresponding $10\sigma$ outlier fractions are
much more similar.
Assessing the overall performance, DC2 appears to be slightly better because
of the improved $\mSBcompare$ in the $i$-band and the reduced $5\sigma$ flux-outliers.
While this DC2-vs-DES comparison is encouraging for \diapipe{},
the processing of simulated DC2 images may be giving somewhat optimistic results
compared to real data.
}

\add{
The SB anomaly is present in both DES and DC2, but still not understood.
Follow-up to DC2 simulations would enable modifying various atmospheric or detector effects
to trace the origin of the anomaly.
Improving DIA is important for studies of transients near cores of bright galaxies.
} 

A similar DIA \add{efficiency} study was done by the Palomar Transient Factory (PTF, \citealp{Frohmaier17}) using $\sim7$ million artificial point sources overlaid on PTF images.
They characterized their real-time detection efficiency as a function  source magnitude,
host galaxy surface brightness, and various observing conditions.
Their efficiencies are mostly parameterized by surface brightness, and thus cannot be directly compared with our DC2 results using SNR. Nevertheless, they report a $\mhalf\simeq 20.3$ in $R$ band for PTF 48 inch class instrument.
As shown in \add{DES, PTF,}
and this DC2 analysis,
the detection efficiency \add{has not} been analytically modelled and
\add{was therefore} determined empirically with fake sources.

\add{DIA} performance depends critically on \add{using} template images with exceptional quality,
in particular a narrower PSF with respect to search images. We created templates using Y1 data,
and found that \add{poor $u$-band seeing} in Y1 (Fig.~\ref{fig:templateseeing})
degraded DIA for $u$ band, where we find a drop in detection efficiency,
as well as biases in photometric flux and uncertainty.
The $u$ band filter transmission is much lower than for the other bands,
and it is unlikely to discover SNIa because they are faint in the UV;
nonetheless, $u$ band is useful for photometric classifiers to distinguish between SNe~Ia and core collapse SNe.

The level of artifact detections from DIA is consistent with
\add{DES} (Table~\ref{tab:DC2_vs_DES}). Machine Learning (ML) methodologies are expected to reduce this contamination significantly,
according to results obtained by several collaborations
\citep{brink_ml_2009, Goldstein_2015,Mahabal_MLZTF_2019, duev2019real}.
We find that the density of artifact detections decreases with increasing PSF size in the search image.
\add{This effect might be caused by the DIA kernel transformation,
which performs better when the search and template PSF difference becomes larger (Liu et al. \textit{in prep.})}. 


%
\add{
\begin{table}[]
    \centering
    \caption{Difference image properties for DC2 and DES. }
\begin{threeparttable}
    \begin{tabular}{c|cc|cc}
    \toprule
                       & DC2-$g$ & DES-$g$\tnote{*} & DC2-$i$ & DES-$i$\tnote{**} \\
    \midrule
    $\mhalf$           & 24.7 & 24.5 & 23.5 & 23.5 \\
    $\SNRhalf$         & 5.57 & 5.61 & 5.84 & 5.36 \\
    $\mSBcompare$ & 22.1 & 22.0 & 19.8 & 21.5 \\
    $\overline{\Bd}$   & 1080 & 520  & 680  & 730 \\
    $f_{5\sigma}(\%)$  & 0.24 & 0.49 & 0.16 & 0.25 \\
    $f_{10\sigma}(\%)$ & 0.09 & 0.09 & 0.04 & 0.06 \\
    \midrule
    \end{tabular}
\begin{tablenotes}\footnotesize
    \item[*]{From DES SN Deep fields, to match DC2-$g$ depth.}
    \item[**]{From DES SN Shallow fields, to match DC2-$i$ depth.}
    \end{tablenotes}
    \end{threeparttable}
    \label{tab:DC2_vs_DES}
\end{table}
} 

\begin{table}[]
    \centering
\begin{threeparttable}
    \caption{$\mSBcompare$\tnote{a} for each DC2 band.} \label{tab:SBRMS}
    \begin{tabular}{c|c|c|c|c|c}
    \toprule
        $u_{\rm{SB}}$ & $g_{\rm{SB}}$ & $r_{\rm{SB}}$ & $i_{\rm{SB}}$ & $z_{\rm{SB}}$ & $y_{\rm{SB}}$ \\
    \midrule
      --\tnote{b}  & 22.1 & 20.5 & 19.8 & 19\tnote{c} & --\tnote{b}  \\
    \midrule
    \end{tabular}
\begin{tablenotes}\footnotesize
    \item[a]{\add{Surface brightness $[\magn/\asec^2]$ where flux-uncertainty scale is $\sim2$}}
    \item[b]{Scale is always $<2$}
    \item[c]{Estimated from extrapolation}
    \end{tablenotes}
    \end{threeparttable}

\end{table}

%

The DC2 baseline cadence is sub-optimal with respect to the recent developments in LSST cadence studies \citep{scolnic_descddfcadence_2018,lochner_descwfdcadence_2018,lochner_impactOS_2021}.
Repeating this DC2 image simulation and analysis on alternative cadences is impractical from both a computational and human-effort perspective. To rigorously evaluate alternative cadences, however, this DC2 analysis demonstrates that the \snana{} simulation can rapidly generate light curve samples that accurately model a full DIA analysis on images. The \snana{} simulation uses meta-data from images and DIA that includes cadence, zero point, PSF, sky noise, detection efficiency vs. SNR, and flux-uncertainty vs. SB. A recommended simulation upgrade is to model catastrophic flux-outliers shown in Table~\ref{tab:outliers}.



\section{Conclusions}
\label{section:conclusions}
In this work, we show results of an integrated Difference Image Analysis pipeline,
built using DESC's \texttt{dia\_pipe} \add{and LSST pipelines},
to analyze simulated images that include SN~Ia light curves.
Using a light curve catalog compiled from \texttt{dia\_pipe} results,
we applied a commonly used SN~Ia standardization method to measure cosmic distances and cosmological parameters. This is the first time that a survey team has carried out such a pixel-to-cosmology test before commissioning operations begin. This analysis is an important stepping stone, enabling monitoring of pipeline performance evolution from survey simulations to real-time analysis during operations.

We have analyzed \area{} of DC2 WFD images using LSST DESC's \texttt{dia\_pipe} pipeline framework for difference-imaging and transient discovery.
\add{The} detection efficiency is ${\sim}100\%$ for point sources with $\rm SNR \geq 8$,
and is $50\%$ efficient for events with $\rm SNR{\sim}5.8$.
\add{
Comparing DC2 and DES in bands with the same search depth,
the difference-image properties are quite similar (Table~\ref{tab:DC2_vs_DES}).
To the extent that the simulated DC2 images are realistic, this comparison
shows that \diapipe{} is already performing at the level of a stage III precursor survey
that was focused on precision measurements of cosmological parameters.
} 

We apply a cosmology analysis using a SALT2+BBC framework, resulting in \NDATACUTS{} SN~Ia light curves (\NLSSTCUTS\ for DC2, \NLOWZCUTS\ for {\LowZName}).
To correct for distance biases in BBC, we used \snana{} to generate a \nameSimData{} simulation of SN~Ia light curves using measured DC2 image properties (PSF, zero point, sky noise) and measured DIA properties (efficiency vs. SNR, flux-uncertainty scale vs. SB).
Both the DC2 and \nameSimData\ simulated samples were used to measure $w$ and $\OM$
from a bias-corrected Hubble diagram; in both cases we recovered the true cosmological and nuisance parameter
values within statistical uncertainties.

We emphasize that the pipeline system is still in active development and may improve by the time LSST starts operations.
%
%
\add{The} pre-commissioning analysis of DC2 is \add{a central contribution} for operational readiness.

\vspace{1cm}

\begin{acknowledgements}
\\
Author contributions are listed below. \\
B.~S\'anchez: Lead design, performed DIA, writing. \\
R.~Kessler: Co-lead project, SNANA simulations and analysis, writing. \\
D.~Scolnic: Lead design and writing. \\
B.~Armstrong: Main \diapipe{} software developer; draft reviewer. \\
R.~Biswas: DC2 simulation; analysis and data preparation. \\
J.~Bogart: DC2 database storage, access and curation. \\
J.~Chiang: Computing infrastructure, DC2 data access and analysis. \\
J.~Cohen-Tanugi: DC2 image processing and data retrieval.  \\
D.~Fouchez: Internal reviewer; suggested paper text and figures edits. \\
P.~Gris: SNWG convener; SNIa discussions. \\
K.~Heitmann: DC2 simulations, paper comments. \\
R.~Hlo\v{z}ek: Early discussions, DC2 analysis software. \\
S.~Jha: Internal reviewer; suggested edits to paper text and figures. \\
H.~Kelly: Computing infrastructure; data processing. \\
S.~Liu: DIA analysis tests and debugging. \\
G.~Narayan: DIA analysis discussion, SNIa discussions \\
B.~Racine: DIA analysis discussion; data visualization. \\
E.~Rykoff: DC2 metadata extraction. \\
M.~Sullivan: TD Working Group convener.  \\
C.~Walter: DIA discussion and comments. \\
M.~Wood-Vasey: Consulted on several DIA aspects; DESC builder. \\

\\
B.S and D.S. are supported by DOE grant DE-SC0010007. D.S. is also supported by DOE grant DE-SC0021962 and the David and Lucile Packard Foundation.
\input{descackstandard}
\\
This work was completed in part with resources provided by the University of Chicago’s Research Computing Center.
This work has gone through DESC internal review process, and main authors would like to explicitly thank Saurabh Jha, Dominique Fouchez, and Bob Armstrong for their comments.
\\
This research has made use of the following \texttt{Python} software packages: \texttt{Astropy} \citep{astropy:2013,astropy:2018}, \texttt{Matplotlib} \citep{hunter2007matplotlib}, \texttt{Pandas} \citep{mckinney2010data}, \texttt{NumPy} \citep{walt2011numpy}, \texttt{Seaborn} \citep{michael_waskom_2014_12710}, \texttt{SciPy} \citep{jones2001scipy}.
\end{acknowledgements}

%
\label{biblio}
\bibliographystyle{aa}
\normalsize
\bibliography{aanda}


\end{document}

%% file: descackstandard.tex
The DESC acknowledges ongoing support from the Institut National de 
Physique Nucl\'eaire et de Physique des Particules in France; the 
Science \& Technology Facilities Council in the United Kingdom; and the
Department of Energy, the National Science Foundation, and the LSST 
Corporation in the United States.  DESC uses resources of the IN2P3 
Computing Center (CC-IN2P3--Lyon/Villeurbanne - France) funded by the 
Centre National de la Recherche Scientifique; the National Energy 
Research Scientific Computing Center, a DOE Office of Science User 
Facility supported by the Office of Science of the U.S.\ Department of
Energy under Contract No.\ DE-AC02-05CH11231; STFC DiRAC HPC Facilities, 
funded by UK BIS National E-infrastructure capital grants; and the UK 
particle physics grid, supported by the GridPP Collaboration.  This 
work was performed in part under DOE Contract DE-AC02-76SF00515.